%% file: PionHBTpaper.tex
\documentclass[twocolumn,secnumarabic,tightenlines,amssymb,amsmath,nobibnotes,aps,prc,superscriptaddress]{revtex4}
\usepackage{bm}
\usepackage{vmargin}
\usepackage{graphicx}
\expandafter\ifx\csname package@font\endcsname\relax\else
 \expandafter\expandafter
 \expandafter\usepackage
 \expandafter\expandafter
 \expandafter{\csname package@font\endcsname}
\fi

\begin{document}

\title{Pion interferometry in Au+Au collisions at $\sqrt{s_{NN}}$ = 200 GeV}
\input{sci-jan04}

\begin{abstract}
We present a systematic analysis of two-pion interferometry in
Au+Au collisions at $\sqrt{s_{NN}}$ = 200 GeV using the STAR
detector at RHIC. We extract the HBT radii and study their
multiplicity, transverse momentum, and azimuthal angle dependence.
The Gaussianess of the correlation function is studied. Estimates
of the geometrical and dynamical structure of the freeze-out
source are extracted by fits with blast wave parameterizations.
The expansion of the source and its relation with the initial
energy density distribution is studied.

\end{abstract}

\maketitle

\section{Introduction}\label{Int}

In the late 1950's two particle intensity interferometry (HBT) was
proposed and developed by the astronomers Hanbury-Brown and Twiss
to measure the angular size of distant stars~\cite{HANBU56}. In
1960, Goldhaber \textit{et al.} applied this technique to particle
physics to study the angular distribution of identical pion pairs
in $p\bar{p}$ annihilations \cite{GOLDH60}. They observed an
enhancement of pairs at small relative momenta that was explained
in terms of the symmetrization of the two-pion wave function.

In ultra-relativistic heavy ion collisions, where a quark gluon
plasma (QGP) is expected to be formed, HBT is a useful tool to
study the space-time geometry of the particle-emitting source
\cite{BAUER92,HEINZ99}. It also contains dynamical information
that can be explored by studying the transverse momentum
dependence of the apparent source size \cite{PRATT84,MAKHL88}. In
non-central collisions, information on the anisotropic shape of
the pion-emitting region at kinetic freeze-out can be extracted by
measuring two-pion correlation functions as a function of the
emission angle with respect the reaction plane, see, for example,
\cite{VOLOS96,VOLOS96b,HEINZ02b}.

Experimentally, two-particle correlations are studied by
constructing the correlation function as \cite{HEINZ99}:
\begin{equation}\label{ExpCF}
C(\vec{q}) = \frac{A(\vec{q})}{B(\vec{q})}.
\end{equation}
Here $A(\vec{q})$ is the pair distribution in momentum difference
$\vec{q}$ = $\vec{p_1} - \vec{p_2}$ for pairs of particles from
the same event, and $B(\vec{q})$ is the corresponding distribution
for pairs of particles from different events. To good
approximation this ratio is sensitive to the spatial extent of the
emitting region and insensitive to the single particle momentum
distribution, acceptance, and efficiency effects \cite{HEINZ99}.

At the Relativistic Heavy Ion Collider (RHIC), identical-pion HBT
studies at $\sqrt{s_{NN}}$ = 130 GeV \cite{ADLER01,ADCOX02} lead
to an apparent source size qualitatively similar to measurements
at lower energies. In contrast to predictions of larger sources
based on QGP formation \cite{RISCH96,RISCH96b}, no long emission
duration is seen. The extracted parameters do not agree with
predictions of hydrodynamic models that, on the other hand,
describe reasonably well the momentum-space structure of the
emitting source and elliptic flow \cite{KOLB03b}. This ``HBT
puzzle" could be related to the fact that the extracted timescales
are smaller than those predicted by the hydrodynamical model
\cite{KOLB03b}. More sophisticated approaches such as 3D
hydrodynamical calculations \cite{HIRAN04} or multi-stage models
\cite{ZWLIN02} also cannot describe simultaneously the geometry
and the dynamic of the system \cite{NOTE1}.

Further detail may be obtained from non-central collisions, where
the initial anisotropic collision geometry has an almond shape
with its longer axis perpendicular to the reaction plane. This
generates greater transverse pressure gradients in the reaction
plane than perpendicular to it. This leads to preferential
in-plane expansion \cite{OLLIT92,KOLB00,ADLER01b,VOLOS03} which
diminishes the initial anisotropy as the source evolves. Thus, the
source shape at freeze-out should be sensitive to the evolution of
the pressure and the system lifetime. Hydrodynamic calculations
\cite{KOLB03} predict that the source may still be out-of-plane
extended after hydrodynamic evolution. However a subsequent
rescattering phase tends to make the source in-plane extended
\cite{TEANE01}. Therefore, the experimental freeze-out source
shape might discriminate between different scenarios of the
system's evolution.

In this paper we present results of our systematic studies of
two-pion HBT correlations in Au+Au collisions at $\sqrt{s_{NN}}$ =
200 GeV measured in the STAR (Solenoidal Tracker at RHIC) detector
at RHIC. We describe the analysis procedure in detail and discuss
several issues with importance to HBT such as different ways of
taking the final state Coulomb interaction into account, and the
Gaussianess of the measured correlation function. The paper is
organized as follows: Section \ref{ExpSetup} introduces the
experimental setup as well as the event, particle and pair
selections. In section \ref{AnalysisMethod}, the analysis method
is presented. In section \ref{results}, systematic results are
shown. We discuss these results in section \ref{Discussion}, where
the centrality dependence of the transverse mass $m_T$ dependence
of the HBT parameters is investigated, the extracted parameters
from a fit to a blast wave parametrization are discussed in detail
and the expansion of the source is studied. We summarize and
conclude in section \ref{Concl}. Extended details about the
analysis method, the results and the discussion can be found in
reference \cite{LOPEZ04}.

\section{Experimental setup, event and particle selection}\label{ExpSetup}
\subsection{STAR detector}\label{STAR}

The STAR detector is an azimuthally symmetric, large acceptance,
solenoidal detector. The subsystems relevant for this analysis are
a large Time Projection Chamber (TPC) located in a 0.5 Tesla
solenoidal magnet, two zero-degree calorimeters (ZDCs) that detect
spectator neutrons from the collision, and a central trigger
barrel (CTB) that measures charged particle multiplicity. The
latter two subsystems were used for online triggering only.

The TPC \cite{ANDER03} is the primary STAR detector and the only
detector used for the event reconstruction of the analysis
presented here. It is 4.2 m long and covers the pseudorapidity
region $|\eta| <$ 1.8 with full azimuthal coverage ($-\pi < \phi <
\pi$). It is a gas chamber, with inner and outer radii of 50 and
200 cm respectively, in a uniform electric field. The particles
passing through the gas release secondary electrons that drift to
the readout end caps at both ends of the chamber. The readout
system is based on multiwire proportional chambers, with readout
pads. There are 45 pad-rows between the inner and the outer radii
of the TPC. The induced charge from the electrons is shared over
several adjacent pads, so the original track position is
reconstructed to $\sim$~500 $\mu$m precision.

The STAR trigger detectors are the CTB and the ZDCs. In this
analysis two trigger settings were used. Hadronic minimum-bias
that requires a signal above threshold in both ZDCs, and hadronic
central that requires low ZDC signal and high CTB signal.

\subsection{Event selection and binning}\label{selection}

For this analysis, we selected events with a collision vertex
position within $\pm$ 25 cm measured along the beam axis from the
center of the TPC. This event \textit{selection} was applied to
all data sets discussed here.

We further binned events by centrality, where the centrality was
characterized according to the measured multiplicity of charged
hadrons with pseudorapidity ($|\eta| <$ 0.5), and here we present
results as a function of centrality bins. The six centrality bins
correspond to 0--5\%, 5--10\%, 10--20\%, 20--30\%, 30--50\%, and
50--80\% of the total hadronic cross-section. A hadronic-central
triggered data set of 1 million events was used only for the first
bin. The other five bins are from a minimum-bias triggered data
set of 1.7 million events.

Within each centrality bin, in order to form the background pairs
for the correlation function (see section \ref{CF}), we only mixed
``similar" events. In this analysis, ``similar" events have
primary vertex relative $z$ position within 5 cm, multiplicities
within the same centrality bin described above, and, for the
azimuthally-sensitive analysis, estimated reaction plane
orientations within $20^\circ$.

\subsection{Particle selection}

We selected tracks in the rapidity region $|y| <$ 0.5. Particle
identification was done by correlating the specific ionization of
the particles in the gas of TPC with their measured momentum. The
energy lost by a particle as it travels through a gas depends on
the velocity $\beta$ at which it travels and it is described by
the Bethe-Bloch formula \cite{WBLUM93}.

For a given momentum, each particle mass will have a different
velocity and a different \textit{dE/dx} as it goes through the gas
of the TPC. For this analysis pions were selected by requiring the
specific ionization to be within 2 standard deviations
(experimentally determined as a function of the particle momentum
and event multiplicity) from the Bethe-Bloch value for pions. To
help remove kaons that could satisfy this condition, particles
were also required to be farther than 2 standard deviations from
the value for kaons. There is a small contamination of electrons
in the low momentum region, $p < 400$ MeV/c. Its effect was
studied with different cuts and found to be unimportant.

To reduce contributions from non-primary (decay) pions, we applied
a cut of 3 cm to each track on the distance of closest approach of
the extrapolated track to the primary vertex.

In our previous HBT analysis \cite{ADLER01}, tracks were divided
in different bins according to their transverse momentum, $p_T$,
and only particles within a given bin were used to form
correlation functions. In this analysis no such $p_T$ binning was
applied. The $p_T$ range was set by limitations in the
reconstruction of pions in the TPC, by the fact that we remove the
kaon band and by the momentum pair cut described below and only
tracks with $150 < p_T < 800$ MeV/c were accepted.

\subsection{Pair cuts}\label{PairCuts}

In this section we describe pair cuts and binning. The first two
cuts discussed are intended to remove the effects of two track
reconstruction defects that are important to HBT: split tracks
(one single particle reconstructed as two tracks) and merged
tracks (two particles with similar momenta reconstructed as one
track).

\subsubsection{Split tracks}

\begin{figure}
\includegraphics[width=0.40\textwidth]{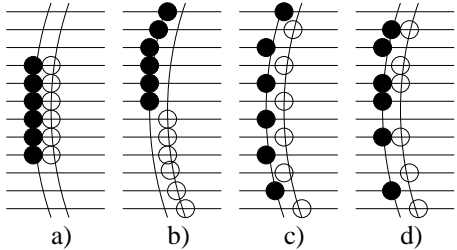}
\caption{Distribution of same number of hits in two tracks for
four possible cases. Closed circles are hits assigned to one
track, open circles are assigned to the other. a$)$ SL = --0.5
(clearly two tracks) b$)$ SL = 1 (possible split track) c$)$ SL =
1 (possible split track) d$)$ SL = 0.08 (likely two
tracks).}\label{SplittingSketch}
\end{figure}

Track splitting causes an enhancement of pairs at low relative
momentum $q$. This false enhancement is created by single tracks
reconstructed as two with similar momenta. In order to remove
split tracks we compare the location of the hits for each track in
the pair along the pad-rows in the TPC and assign a quantity to
each pair, called \textit{Splitting Level} (SL), calculated as
follows:
\begin{eqnarray}
\lefteqn{\textrm{SL} \equiv \frac{\sum_{i} S_i}{\textrm{Nhits}_1
+ \textrm{Nhits}_2}~~~\textrm{where}~~~\textrm{S}_i = {}}\\
& & \Bigg\{%
\begin{array}{ll}
 +1 & \textrm{one~track~leaves~a~hit~on~pad-row}\nonumber\\
 -1 & \textrm{both~tracks~leave~a~hit~on~pad-row}\nonumber\\
 0 & \textrm{neither~track~leaves~a~hit~on~pad-row},\nonumber\\
\end{array}
\end{eqnarray}
where $i$ is the pad-row number, and $\textrm{Nhits}_1$ and
$\textrm{Nhits}_2$ are the total number of hits associated to each
track in the pair. If only one track has a hit in a pad-row +1 is
added to the running quantity, if both tracks have a hit in the
same pad-row, a sign of separate tracks, --1 is added to this
quantity. After the sum is done, it is divided by the sum of hits
in both tracks, this normalizes SL to a value between --0.5 (both
tracks have hits in exactly same pad-rows) and 1.0 (tracks do not
have any hit in same pad-row). Figure \ref{SplittingSketch} shows
four different cases for the same number of total hits: in case a)
two different tracks with SL = --0.5, in b) and c) two different
cases of possible split tracks with SL = 1, and in d) two
different tracks with SL = 0.08.

\begin{figure}
\includegraphics[width=0.50\textwidth]{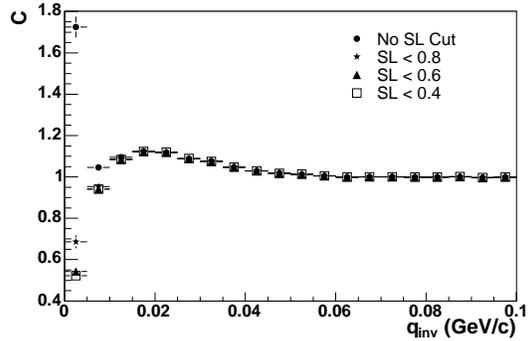}
\caption{1D correlation function for different values of SL
(anti-splitting cut). The cut applied in the analysis is SL $<$
0.6. The horizontal lines indicate the bin
width.}\label{QinvCFvsQual}
\end{figure}

We required every pair to have SL smaller than a certain value.
This value was determined from the 1-dimensional correlation
functions as a function of the relative momentum of the pair
$q_{\textrm{inv}}$, for different values of SL; some of them are
shown in Fig.~\ref{QinvCFvsQual}. The relative momentum of the
pair is defined as $q_{\textrm{inv}} = \sqrt{(q^0)^2 -
|\vec{q}|^2}$ where $q^0$ and $\vec{q}$ are the components of the
four-vector momentum difference. We observe that when making this
cut more restrictive (reducing the maximum allowed value for SL)
the enhancement is reduced until we reach SL = 0.6 when the
correlation function becomes stable and does not change for lower
values of SL. Therefore, all the pairs entering the correlation
functions were required to have SL $<$ 0.6. Cutting at this value
is also supported by simulation studies. While naturally track
splitting can only give rise to false pairs, our SL cut also
removes some real pairs which happen to satisfy the cut.
Therefore, we apply the SL cut to both, ``real" and ``mixed"
pairs, numerator and denominator of $C(\vec{q})$,
Eq.~(\ref{ExpCF}).

\begin{figure}
\includegraphics[width=0.50\textwidth]{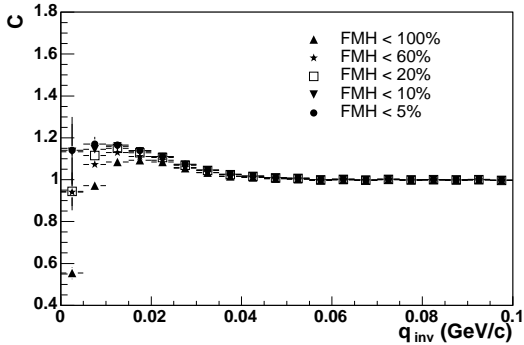}
\caption{1D correlation function for different values of the
maximum fraction of merged hits allowed. Cut applied in the
analysis is fraction of merged hits (FMH) $<$ 10\%. The horizontal
lines indicate the bin width.} \label{QinvCFvsMFMH}
\end{figure}

\subsubsection{Merged tracks}

Once we have removed split tracks we can study the effects of two
particles reconstructed as one track. These merged tracks cause a
reduction of pairs at low relative momentum since the particles
that have higher probability of being merged are those with
similar momenta. To eliminate the effect of track merging, we
required that all pairs entering numerator and denominator of the
correlation function had a fraction of merged hits no larger than
10\%. Two hits are considered merged if the probability of
separating them is less than 99\%. From simulation and data
studies, this minimum separation was determined to be 5 mm. By
applying this cut to ``real" and ``mixed" pairs, we introduce in
the denominator the effect that merged tracks have in the
numerator: a reduction of low \textit{q} pairs. Note that tracks
from different events will originate from primary vertices at
different positions along the beam direction. Thus, even two
tracks with \textit{identical} momenta, which would surely be
merged if they originated from the same event, may not be
considered as a merged track when they formed a ``mixed" pair if
we would not account for the different primary vertex position.
Our procedure is to calculate (using a helix model) the pad-row
hit positions of each track, assuming that the track originated at
the center of the TPC \cite{LOPEZ04}. These calculated hit
positions are used in the merging cut procedure described above.
By applying this cut to the numerator, we would remove ``real"
pairs that satisfy the cut. This would reduce the HBT fit
parameters for a correlation function that is not completely
Gaussian, and needs to be taken into account as will be described
in the next section.

To determine the maximum fraction of merged hits allowed we
proceed as we did for the anti-splitting cut. Figure
\ref{QinvCFvsMFMH} shows the 1-dimensional correlation functions
as a function of $q_{\textrm{inv}}$, for different values of the
maximum fraction of merged hits allowed. By requiring the fraction
of merged hits to be less than 10\% for every pair entering the
correlation function, the effect of merged tracks in the
correlation function was almost completely removed as will be
discussed in section \ref{AnalysisMethod}.

\subsubsection{$k_T$ cut and pair binning}

As already mentioned, no explicit $p_T$ cut was applied to single
tracks beyond the requirement for clean PID. However, in addition
to the two cuts already described, pairs were required to have an
average transverse momentum ($k_T = (|\vec{p}_{1T}| +
|\vec{p}_{2T}|)/2$) between 150 and 600 MeV/c. No difference was
observed between the extracted HBT parameters when applying
equivalent $p_T$ or $k_T$ cuts. However statistics improved when
using the latter cut, as two pions from different $p_T$ bins will
be used in a $k_T$-cut analysis, but not in a $p_T$-cut analysis.

Pairs were then binned by $k_T$ in 4 bins that correspond to
[150,250] MeV/c, [250,350] MeV/c, [350,450] MeV/c, and [450,600]
MeV/c. Here the results are presented as a function of the average
$k_T$ (or $m_T = \sqrt{k_T^2 + m_{\pi}^2}$) in each of those bins.

In addition, in the azimuthally-sensitive HBT analysis, to observe
the particle source from a series of angles, pairs were also
binned according to the angle $\Phi = \phi_{\textrm{pair}} -
\Psi_{2}$, where $\phi_{\textrm{pair}}$ is the azimuthal angle of
the pair transverse momentum $\vec{k}_{T}$ and $\Psi_{2}$ is the
second-order event plane azimuthal angle. The first-order event
plane angle is not reconstructible with the STAR detector
configuration for this analysis. Because we use the second-order
reaction plane, $\Phi$ is only defined in the range [0,$\pi$].

\section{Analysis method}\label{AnalysisMethod}
\subsection{Construction of correlation function}\label{CF}
The two-particle correlation function between identical bosons
with momenta $\vec{p}_1$ and $\vec{p}_2$ is defined in
Eq.~(\ref{ExpCF}). As already mentioned, $A(\vec{q})$ is the
measured distribution of the momentum difference for pairs of
particles from the same event and $B(\vec{q})$ is obtained by
mixing particles in separate events \cite{KOPYL77} and represents
the product of single particle probabilities. Each particle in one
event is mixed with all the particles in a collection of events
which in our case consists of 20 events. As discussed before,
events in a given collection have primary vertex $z$ position
within 5 cm, multiplicities within 5 to 30\% of each other, and,
for the azimuthally-sensitive analysis, estimated reaction plane
orientations within $20^\circ$.

\subsection{Pratt-Bertsch parametrization}\label{BP}

In order to probe length scales differentially in beam and
transverse directions, the relative momentum $\vec{q}$ is usually
decomposed in the Pratt-Bertsch (or ``out-side-long") convention
\cite{BERTS98,PRATT86,CHAPMA95}. In this parametrization the
relative momentum vector of the pair $\vec{q}$ is decomposed into
a longitudinal direction along the beam axis, $q_{l}$, an outward
direction parallel to the pair transverse momentum, $q_{o}$, and a
sideward direction perpendicular to those two, $q_{s}$.

We choose as the reference frame, the longitudinal comoving system
(LCMS) frame of the pair, in which the longitudinal component of
the pair velocity vanishes. At midrapidity, in the LCMS frame, and
with knowledge of the second-order but not the first-order
reaction plane, the correlation function is usually parameterized
by a 3-dimensional Gaussian in the relative momentum components as
\cite{HEINZ02b}:
\begin{equation}\label{3DCF}
C(\vec{q}) = 1 + \lambda e^{-q_o^2R_o^2 - q_s^2R_s^2 - q_l^2R_l^2
- 2q_oq_sR_{os}^2}.
\end{equation}
For an azimuthally integrated analysis, the correlation function
is symmetric under $q_s \rightarrow -q_s$ and $R_{os}^2=0$.

In principle, the possibility that the emission of particles is
neither perfectly chaotic nor completely coherent can be taken
into account by adding the parameter $\lambda$ to the correlation
function, which, in general, depends on $k_T$. This $\lambda$
parameter should be unity for a fully chaotic source and smaller
than unity for a source with partially coherent particle emission.
In the analysis presented here we have assumed completely chaotic
emission \cite{ADAMS03e} and attribute the deviations from
$C(q=0,k)=2$ to contribution from pions coming from long-lived
resonances, and misidentified particles, such as electrons.

While for the azimuthally integrated analysis the sign of the
$\vec{q}$ components is arbitrary, in the azimuthally-sensitive
analysis, the sign of $R_{os}^2$ is important because it tells us
the azimuthal direction of the emitted particles, so the signs of
$q_o$ and $q_s$ are kept and particles in every pair are ordered
such that $q_l > 0$.

References \cite{WIEDE98,WIEDE99} give a detailed description of
the relation between the HBT radius parameters ($R_o^2$, $R_s^2$,
$R_l^2$ and $R_{os}^2$) and the space-time geometry of the final
freeze-out stage.

\subsection{Fourier components}

For a boost-invariant system, the $\Phi$ dependence of the HBT
radii of Eq.~(\ref{3DCF}) are \cite{HEINZ02b}:

\begin{eqnarray}\label{asRadii}
\lefteqn{ R_{\mu}^2(k_T,\Phi) = R_{\mu,0}^2(k_T) + {} }\nonumber\\
& & {} 2\sum_{n=2,4,6\cdots}R_{\mu,n}^2(k_T)\cos(n\Phi)~~~(\mu=o,s,l){}\nonumber\\
\lefteqn{R_{\mu}^2(k_T,\Phi)={} }\nonumber\\
& & {}
2\sum_{n=2,4,6\cdots}R_{\mu,n}^2(k_T)\sin(n\Phi)~~~(\mu=os),
\end{eqnarray}
where $R_{\mu,n}(k_T)$ are the $n^{\textrm{th}}$ order Fourier
coefficients for the $\mu$ radius. These coefficients, that are
$\Phi$ independent, can be calculated as:

\begin{equation}\label{FouCoef}
R_{\mu,n}^2 (k_T) = \Bigg\{%
\begin{array}{ll}
  \langle R_{\mu}^2(k_T,\Phi)\cos(n\Phi)\rangle & (\mu=o,s,l)\\
  \langle R_{\mu}^2(k_T,\Phi)\sin(n\Phi)\rangle & (\mu=os).
\end{array}
\end{equation}

As we will show, the $0^{\textrm{th}}$ order Fourier coefficients
correspond to the extracted HBT radii in an azimuthally integrated
analysis. In this analysis we found that Fourier coefficients
above $2^{\textrm{nd}}$ order are consistent with 0.

\subsection{Coulomb interaction and fitting procedures}\label{CC}

\begin{figure}
\includegraphics[width=0.50\textwidth]{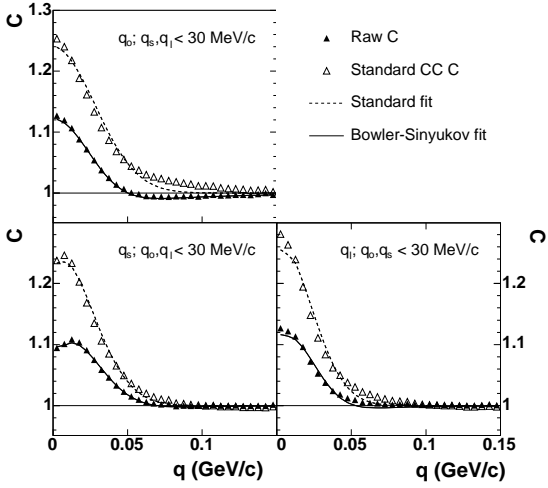}
\caption{Projections of the 3 dimensional correlation function and
corresponding fits for negative pions from the 0--5\% most central
events and $k_T$ = [150,250] MeV/c according to the
\textit{standard} and \textit{Bowler-Sinyukov}
procedures.}\label{3Dprojections}
\end{figure}

Equation (\ref{3DCF}) applies only if the sole cause of
correlation is quantum statistics and the correlation function is
Gaussian. We come to this second point in section \ref{Gaussian}.
In our case, significant Coulomb effects must also be accounted
for (strong interactions are within reasonable limit here
\cite{WIEDE99}). This Coulomb interaction between pairs, repulsive
for like-sign particles, causes a reduction in the number of real
pairs at low $q$ reducing the experimental correlation function as
seen in Fig.~\ref{3Dprojections}.

\subsubsection{\textit{Standard} procedure}

Three different procedures can be applied in order to take this
interaction into account. One procedure that was used in our
analysis at $\sqrt{s{NN}}$ = 130 GeV \cite{ADLER01} as well as by
previous experiments, consists of fitting the correlation function
to

\begin{eqnarray}\label{StandFit}
\lefteqn{C(q_o, q_s, q_l) = \frac{A(\vec{q})}{B(\vec{q})}
= K_{\textrm{coul}}(q_{\textrm{inv}})\times}\\
& & (1 + \lambda e^{-q_o^2R_o^2 - q_s^2R_s^2 - q_l^2R_l^2 -
2q_oq_sR_{os}^2}),\nonumber
\end{eqnarray}
normalized to unity at large $\vec{q}$, where $K_{\textrm{coul}}$
is the squared Coulomb wave-function integrated over the whole
source, which in our case is a spherical Gaussian source of 5 fm
radius. The effect on the final results of changing the radius of
the spherical Gaussian source to calculate $K_{\textrm{coul}}$ was
studied and found to be within reasonable limits. Traditionally,
Eq.~(\ref{StandFit}) has been expressed as
\begin{eqnarray}\label{StandFit2}
\lefteqn{C'(q_o, q_s, q_l) = \frac{A(\vec{q})}{B(\vec{q})
K_{\textrm{coul}}(q_{\textrm{inv}})} }\\
& & = 1 + \lambda e^{-q_o^2R_o^2 - q_s^2R_s^2 - q_l^2R_l^2 -
2q_oq_sR_{os}^2},\nonumber
\end{eqnarray}
and this new correlation function was called the \textit{Coulomb
corrected correlation function} since we introduce in the
denominator a Coulomb factor with which we try to compensate the
Coulomb interaction in the numerator. We call this
\textit{standard} procedure. However, this procedure overcorrects
the correlation function since it assumes that all pairs in the
background are primary pairs and need to be corrected
\cite{PRATT86b}.

\subsubsection{\textit{Dilution} procedure}

In a second procedure, inspired by the previous procedure and
implemented before by the E802 collaboration \cite{AHLE02}, the
Coulomb term is ``diluted" according to the fraction of pairs that
Coulomb interact as:
\begin{equation}\label{KcoulPrime}
K_{\textrm{coul}}'(q_{\textrm{inv}}) = 1 + f
(K_{\textrm{coul}}(q_{\textrm{inv}}) - 1),
\end{equation}
where $f$ has a value between 0 (no Coulomb weighting) and 1
(\textit{standard} weight). The correlation function in this
procedure is fitted to:
\begin{eqnarray}\label{DilutFit}
\lefteqn{C(q_o, q_s, q_l) = \frac{A(\vec{q})}{B(\vec{q})}= K_{\textrm{coul}}'(q_{\textrm{inv}})\times}\\
&& (1 + \lambda e^{-q_o^2R_o^2 - q_s^2R_s^2 - q_l^2R_l^2 -
2q_oq_sR_{os}^2}),\nonumber
\end{eqnarray}
normalized to unity at large $\vec{q}$. We call this the
\textit{dilution} procedure. A reasonable assumption is to take $f
= \lambda$ assuming that $\lambda$ is the fraction of primary
pions. This increases $R_o$ by 10--15\% and has a very small
effect on $R_s$ and $R_l$ as seen in Fig.~\ref{StDlBwCoulomb}.
$\lambda$ decreases by 10--15\%.

\subsubsection{\textit{Bowler-Sinyukov} procedure}

An advantage of the previous two techniques is that after
``correcting" for Coulomb effects, one winds up with a correlation
function which may be fit with a simple Gaussian form. However if
there exists more than one source of interaction, it is not valid
to ``correct" one way. For example, it is, in fact, the same pion
pairs which Coulomb interact as which show quantum enhancement.
This leads to a change in the expected form of the correlation
function if not all particles participate in the interaction
(\textrm{i.e.}, $\lambda\neq 0$) \cite{NOTE2}. If $\lambda = 1$,
all three methods are equivalent.

\begin{figure}
\includegraphics[width=0.50\textwidth]{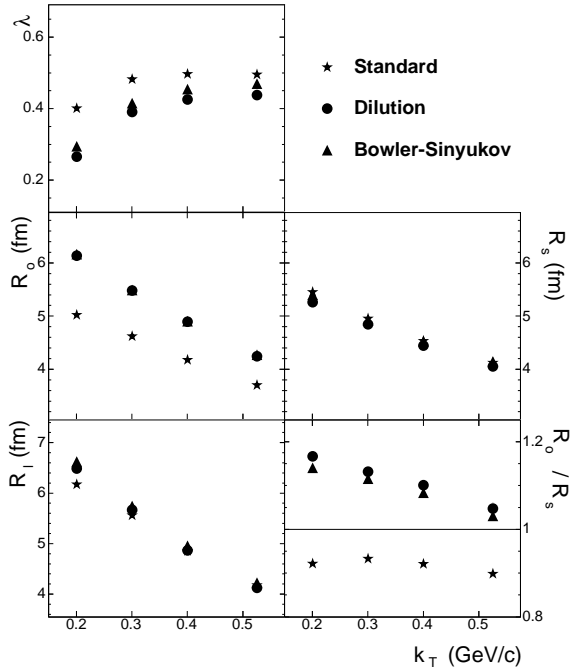}
\caption{HBT parameters for the three possible fitting procedures
to the correlation functions described in this paper depending on
how Coulomb interaction is taken into account from the 0--5\% most
central events. Error bars contain only statistical
uncertainties.} \label{StDlBwCoulomb}
\end{figure}

In this analysis, we have implemented a new procedure, first
suggested by Bowler \cite{BOWL91} and Sinyukov \textit{et al}.
\cite{SINY98}, and recently advocated by the CERES collaboration
\cite{ADAMO03}, in which only pairs with Bose-Einstein interaction
are considered to Coulomb interact. The correlation function in
this procedure is fitted to:
\begin{eqnarray}\label{BowlerFit}
\lefteqn{C(q_o, q_s, q_l) = \frac{A(\vec{q})}{B(\vec{q})}=
(1-\lambda) +}\\
& & \lambda K_{\textrm{coul}}(q_{\textrm{inv}})(1 + e^{-q_o^2R_o^2
- q_s^2R_s^2 - q_l^2R_l^2 - 2q_oq_sR_{os}^2}),\nonumber
\end{eqnarray}
normalized to unity at large $\vec{q}$, where
$K_{\textrm{coul}}(q_{\textrm{inv}})$ is the same as in the
\textit{standard} procedure. The first term on the right-hand side
of Eq.~(\ref{BowlerFit}) accounts for the pairs that do not
interact and the second term for the pairs that (Coulomb and
Bose-Einstein) interact. We call this \textit{Bowler-Sinyukov}
procedure. It has a similar effect on the HBT parameters as the
\textit{dilution} procedure as seen in Fig.~\ref{StDlBwCoulomb}. A
similar procedure has been recently implemented by the Phobos
collaboration \cite{BACK04}. In this procedure only pairs which
are close in the pair center of mass frame are considered to
Coulomb interact.

It is worth mentioning that the parameters $\lambda$ and $R_o$,
and consequently the ratio $R_o/R_s$, extracted using the
\textit{standard} procedure here are smaller than the parameters
obtained in our previous analysis \cite{ADLER01}. This is
explained by a different particle selection. In the analysis
presented here, the contribution from non-primary pions is larger
than in the previous analysis, leading to smaller $\lambda$ and
$R_o$ when using that procedure. However, the parameters obtained
when applying the \textit{Bowler-Sinyukov} procedure are almost
not affected by the contribution from non-primary pions.

\subsubsection{Comparison of methods}

Figure \ref{3Dprojections} shows the projections of the
3-dimensional correlation function according to the Pratt-Bertsch
parametrization described in section \ref{BP} for an azimuthally
integrated analysis. The closed symbols represent the correlation
function and the open symbols the \textit{Coulomb corrected
correlation function} according to the \textit{standard}
procedure. The lines are fits to the data, the dashed line is the
\textit{standard} fit to the \textit{Coulomb corrected correlation
function}, and the continuous line the \textit{Bowler-Sinyukov}
fit to the uncorrected correlation function. The extracted
parameters from both fits are the parameters for the lowest
$\langle k_T\rangle$ in Fig.~\ref{StDlBwCoulomb}.

\begin{figure}
\includegraphics[width=0.50\textwidth]{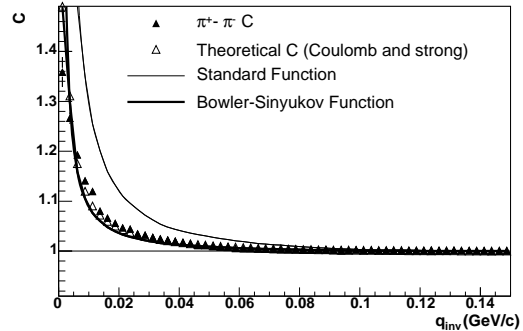}
\caption{1D correlation function for $\pi^+\pi^-$ compared to
Standard, Bowler-Sinyukov functions and a theoretical calculation
that includes Coulomb and strong interactions.}\label{PipPimCFs}
\end{figure}

As a consistency check for the \textit{Bowler-Sinyukov} procedure,
we calculated the $\pi^+\pi^-$ correlation function, dominated by
Coulomb interaction, and compare to different calculations. In
Fig.~\ref{PipPimCFs} lines indicate the \textit{standard}
($K_{c\textrm{oul}}(q_{\textrm{inv}})$) and
\textit{Bowler-Sinyukov} ($(1 - \lambda) + \lambda
K_{\textrm{coul}}(q_{\textrm{inv}})$) Coulomb functions where
$\lambda$ was extracted from the fit to the 3D like-sign
correlation function. This latter $\lambda$ is the same $\lambda$
as \textit{dilution} for unlike sign pions and takes into account
the percentage of primary pions through $\lambda$. Clearly, the
\textit{Bowler-Sinyukov} function (thick line) better reproduces
the data (closed symbols) than the \textit{standard} function
(thin line). The small discrepancy between the
\textit{Bowler-Sinyukov} function and the data disappears when
strong interaction (negligible for like-sign pions) is added to
the \textit{Bowler-Sinyukov} function as shown by the theoretical
calculation \cite{LEDNI82} (open symbols). Between identical
pions, there is a repulsive s-wave interaction for the isospin I=2
system \cite{BOWLE88}. However, the range of this interaction is
estimated to be $\sim$~0.2 fm, while the characteristic separation
between pions in heavy ions collisions is $\sim$~5 fm. Also, there
are no doubly charged mesonic resonances that could decay into
same charged pions that would strongly interact. For these
reasons, the strong interaction will be ignored for like sign
particles.

\subsubsection{Coulomb interaction with the source}

The Coulomb interaction between the outgoing charged pions and the
residual positive charge in the source is negligible
\cite{WZAJC84,HBARZ99}. This is confirmed by the good agreement
observed between the parameters extracted from $\pi^+\pi^+$ and
$\pi^-\pi^-$ correlation functions as shown later in this paper,
Fig.~\ref{HBTpar_vsmT_C1}.

\subsection{Momentum resolution correction}\label{MomResCorr}
\begin{figure}
\includegraphics[width=0.45\textwidth]{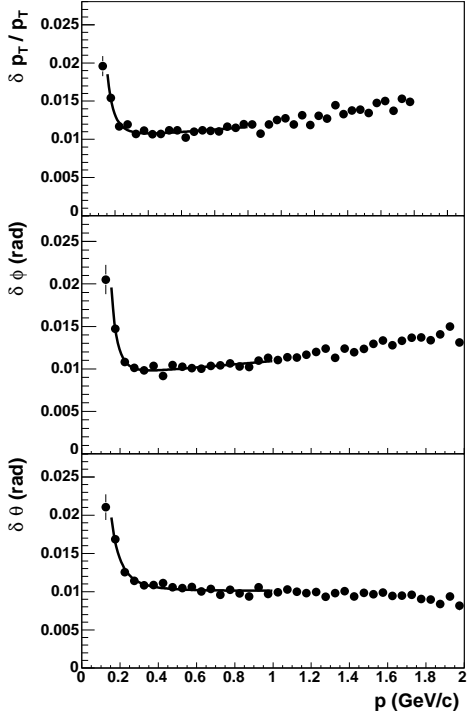}
\caption{Momentum resolution for pions at midrapidity expressed by
the widths $\delta p_T/p_T$, $\delta\varphi$ and $\delta\theta$ as
a function of $p$.}\label{MomResWidths}
\end{figure}

The limited single-particle momentum resolution induces broadening
of the correlation function and thus systematic underestimation of
the HBT parameters. To determine the magnitude of this effect we
need to know the momentum resolution for the particles under
consideration. We estimate our single-particle momentum resolution
by embedding simulated particles into real events at the TPC pixel
level and comparing the extracted and input momenta. Figure
\ref{MomResWidths} shows the RMS spreads as a function of
$|\vec{p}|$ in $p_T$ and angles $\phi$ and $\theta$, where
$\theta$ is the angle between the momentum of the particle and the
beam axis and $\phi$ is the azimuthal angle of the particle. We
see that the resolution in $p_T$, given by $\delta p_T/p_T$ (top
panel of Fig.~\ref{MomResWidths}), has a width of about 1\% for
the momentum range under consideration.

To account for this limited momentum resolution, a correction,
$K_{\textrm{momentum}}(\vec{q})$, is applied to each measured
correlation function:
\begin{equation}
C(\vec{q}_{\textrm{}}) =
\frac{A(\vec{p}_{1\textrm{meas}},\vec{p}_{2\textrm{meas}})}{B(\vec{p}_{1\textrm{meas}},\vec{p}_{2\textrm{meas}})}
K_{\textrm{momentum}}(\vec{q}).
\end{equation}
The correction factor is calculated from the single-particle
momentum resolution as follows:
\begin{equation}
K_{\textrm{momentum}} (\vec{q}) =
\frac{C(\vec{q}_{\textrm{ideal}})}{C(\vec{q}_{\textrm{smear}})}
\nonumber\\
=
\frac{\frac{A(\vec{p}_{1\textrm{ideal}},\vec{p}_{2\textrm{ideal}})}{B(\vec{p}_{1\textrm{ideal}},\vec{p}_{2\textrm{ideal}})}}
{\frac{A(\vec{p}_{1\textrm{smear}},\vec{p}_{2\textrm{smear}})}{B(\vec{p}_{1\textrm{smear}},\vec{p}_{2\textrm{smear}})}},
\end{equation}
where the \textit{ideal} and \textit{smear} correlation function
are formed as follows. Numerator and denominator of the
\textit{ideal} correlation function are formed by pairs of pions
from different events. Each pair in the numerator is weighted,
according with the \textit{Bowler-Sinyukov} function, by:
\begin{eqnarray}\label{weight}
\lefteqn{weight = (1 - \lambda) + \lambda K_{\textrm{coul}}(q_{\textrm{inv}})\times {}} \nonumber\\
& & (1 + e^{-q_o^2R_o^2 - q_s^2R_s^2 - q_l^2R_l^2 -
2q_oq_sR_{os}^2}),
\end{eqnarray}
where $K_{\textrm{coul}}(q_{\textrm{inv}})$ is the same factor as
described in section \ref{CC} and $R_{os}^2$ = 0 in the
azimuthally integrated analysis. If the measured momentum were the
``real" momentum, this \textit{ideal} correlation function would
be the ``real" correlation function. However this is not the case,
so we calculate a \textit{smeared} correlation function for which
numerator and denominator are also formed by pairs of pions from
different events but their momenta have been smeared according to
the extracted momentum resolution. Pairs in the numerator are also
weighted by the $weight$ given by (\ref{weight}). This
\textit{smeared} correlation function is to the \textit{ideal}
correlation function, as our ``measured" correlation function is
to the ``real" correlation function, which allows us to calculate
the correction factor.

For the $weight$, certain values for the HBT parameters
($\lambda$, $R_{o}$, $R_{s}$, $R_{l}$ and $R_{os}$) need to be
assumed. Therefore, this procedure is iterative with the following
steps:
\begin{enumerate}
    \item Fit the correlation function without momentum resolution
    correction, and use the extracted HBT parameters for the first
    $weight$.
    \item Construct the momentum resolution corrected correlation
    function.
    \item Fit it according to Eq.~(\ref{BowlerFit}).
    \item If the extracted parameters agree with the parameters used to calculate the $weight$,
    those are the final parameters. If they differ
    from the parameters used, then use these latter extracted parameters
    for the new \textit{weight} and go back to step 2.
\end{enumerate}

Also, to be fully consistent, the Coulomb factor
$K_{\textrm{coul}}(q_{\textrm{inv}})$ (where $q_{\textrm{inv}}$ is
calculated from pairs of pions from different events) used in the
fit to extract the HBT parameters must be modified to account for
momentum resolution as follows:
\begin{eqnarray}
\lefteqn{K_{\textrm{coul}}(q_{\textrm{inv}}) {}} \nonumber\\
& & =
K_{\textrm{coul}}(q_{\textrm{inv,meas}})\frac{K_{\textrm{coul}}(q_{\textrm{inv,ideal}})}{K_{\textrm{coul}}(q_{\textrm{inv,smear}})}\nonumber\\
& & =
\frac{K_{\textrm{coul}}^2(q_{\textrm{inv,meas}})}{K_{\textrm{coul}}(q_{\textrm{inv,smear}})}.
\end{eqnarray}

For this analysis, after two iterations the extracted parameters
were consistent with the input parameters. We also checked that
when convergence is reached, the ``uncorrected" HBT parameters
matched the \textit{smeared} parameters. The correction increases
the HBT radius parameters between 1.0\% for the lowest $k_T$ bin
[150,250] MeV/c and 2.5\% for the highest bin [450,600] MeV/c.

\subsection{$\Phi$-dependent HBT analysis methods}\label{asHBTsection}

The study of HBT radii relative to the reaction plane angle was
performed \cite{ADAMS03c} by extending the analysis techniques as
presented in this section, to account for reaction plane
resolution and small instabilities in the fits.  We discuss these
here.  For the azimuthally-sensitive analysis, each of the four
radii extracted from the Bowler-Sinyukov fit contains an implicit
dependence on the azimuthal angle $\Phi$ between the pion pair and
the reaction plane. Azimuthally-sensitive studies of the HBT radii
$\bigl(R_\mu(\Phi)\bigr)$ \cite{ADAMS03c} also must correct for
finite resolution when estimating the true reaction plane
$\Psi_{\rm rp}$ \cite{HEINZ02b}.  Finite reaction plane resolution
acts to decrease the measured amplitude of the radii oscillations,
similar to its effect on azimuthal particle distributions relative
to the reconstructed event plane $\Psi_2$ (\textrm{i.e.}, elliptic
flow \cite{ACKER01}).  The technique for the resolution
correction, which also corrects for finite $\Phi$-bin width, was
developed extensively in Ref.~\cite{HEINZ02b}.  Here we discuss
briefly how this correction is implemented and the resulting
effect on the HBT radii.

The basic principle behind the correction procedure is that, for a
given ${\vec{q}}$-bin in the numerator $A(\vec{q})$ and
denominator $B(\vec{q})$ of each correlation function, the
measured contents for that ${\vec{q}}$-bin at different $\Phi$ are
modified due to the $\Psi_{\rm rp}$ resolution.  The true angular
dependence of $\Phi$ (for each ${\vec{q}}$-bin) can be extracted
from the measured $\Phi_j$ by performing a Fourier decomposition
of $A(\vec{q})$ and $B(\vec{q})$, which leads to the correction
factors \cite{HEINZ02b}

\begin{eqnarray}\label{corrFactors}
A_{\alpha ,n}^{\Delta}(\vec{q}) & = & A_{\alpha,n}(\vec{q})
\frac{\sin (n \Delta /2)}{n \Delta /2}, \\
A^{\textrm{exp}}_{\alpha ,n}(\vec{q}) & = & A_{\alpha,n}^{\Delta}
(\vec{q})
\bigl\langle\cos\left(n(\Psi_2{-}\Psi_{\textrm{rp}})\right)\bigr\rangle,\nonumber
\end{eqnarray}
where $\alpha$ refers to both {\it cosine} and {\it sine} series,
$\Delta$ is the width of each $\Phi$ bin, and $n$ is the Fourier
component.  The factors $\langle\cos\left(n(\Psi_2{-}\Psi_{\rm
rp})\right)\rangle$ are the well-known correction factors for
event plane resolution, obtained by extracting the anisotropic
flow coefficients $v_n$ from the single particle spectrum
\cite{VOLOS96c,POSKA98}. The same procedure is used to correct the
denominator $B(\vec{q})$.

In the present analysis, only the 2nd-order event plane ($\Psi_2$)
is measured.  Using Eq.~(\ref{corrFactors}), the numerator
$A(\vec{q})$ and denominator $B(\vec{q})$ for each ${\vec{q}}$-bin
at each measured angle $\Phi_j$ can be corrected for both the
effects of angular binning and finite event plane resolution:

\begin{eqnarray}\label{Nq}
\lefteqn{A(\vec{q},\Phi_j)~=~N_{\textbf{exp}}(\vec{q},\Phi_j)~+~~~~~~}\nonumber\\
\lefteqn{2~\sum_{n=1}^{n_{\textrm{bin}}} \zeta_{2}(\Delta)
     \Big[A^{\textrm{exp}}_{c,2}(\vec{q})\cos(2\Phi_j)~+}\nonumber\\
& &
A^{\textrm{exp}}_{s,2}(\vec{q})\sin(2\Phi_j)\Big],~~~~~~~~~~~~~~~~~~~~~~
\end{eqnarray}
with the correction parameter $\zeta_{2}(\Delta)$ given by
\begin{equation}
\label{zeta}
  \zeta_{2}(\Delta)=\frac{\Delta}
  {\sin(\Delta)\langle\cos(2(\Psi_2{-}\Psi_{\rm rp}))\rangle_p} - 1.
\end{equation}
The procedure is model independent; the quantities on the right
hand side of Eq.~(\ref{Nq}) are all measured experimentally.

For each set of $\Phi_j$ histograms, the correction procedure
modifies both the numerators and denominators, and therefore the
correlation functions as well.

\begin{figure}
\includegraphics[width=0.50\textwidth]{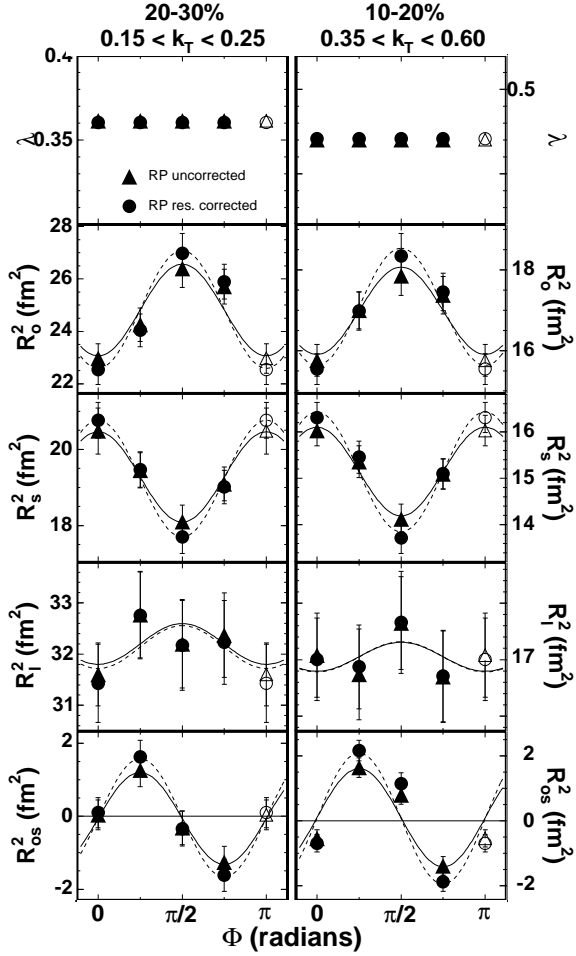}
\caption{Squared HBT radii relative to the reaction plane angle,
without and with the reaction plane resolution applied, for two
different centrality-$k_T$ ranges. The solid lines show allowed
fits to the individual oscillations.}\label{compareResCorr}
\end{figure}

Figure~\ref{compareResCorr} shows the squared HBT radii, obtained
using Eq.~(\ref{asRadii}), as a function of $\Phi$ for two
combinations of centrality and $k_T$. In each case, the
oscillation amplitudes for the three transverse radii increases
after the resolution correction has been applied, while the mean
shows little change.

An additional technique employed in the azimuthally-sensitive HBT
analysis is to use a common $\lambda$ parameter for each
centrality/$k_T$ bin.  This step was undertaken to improve the
quality of the fits by restricting the $\lambda$ parameter, under
the assumption that $\lambda$ should have no implicit $\Phi$
dependence.  For an analysis with four $\Phi$ bins, this
effectively reduces the number of free parameters per
centrality/$k_T$ bin (after normalization) from 20 [(4 radii +
$\lambda$) $\times N_\Phi$] to 17 [(4 radii $\times N_\Phi$) +
$\lambda$].  Since fitting all four correlation functions with a
17-parameter function is arduous, we determined the average
$\lambda$ parameter from the four fits and then re-fit each of the
four correlation functions with $\lambda$ fixed to its average.

\begin{figure}
\includegraphics[width=0.50\textwidth]{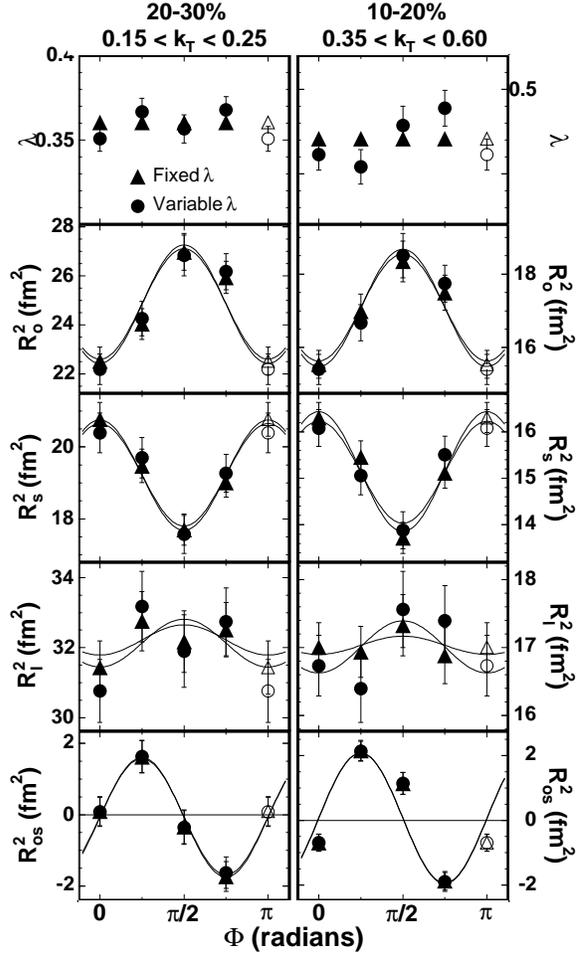}
\caption{Squared HBT radii relative to the reaction plane angle,
for the case where the $\lambda$ parameter is averaged and fixed
in the fit, and the case where $\lambda$ is a free fit parameter
for each $\Phi$. The solid lines show allowed fits to the
individual oscillations.}\label{compareLambda}
\end{figure}

Figure \ref{compareLambda} compares the fit parameters obtained
with and without averaging/fixing $\lambda$, for two
centrality/$k_T$ ranges.  While the individual radii show some
deviations, the resulting Fourier coefficients (which are
represented by the symmetry-constrained drawn in
Fig.~\ref{compareLambda}) are consistent within errors for the two
methods.

\subsection{Systematic uncertainties associated with pair cuts}\label{SystematicErr}

The maximum fraction of merged hits cut described in section
\ref{PairCuts} introduces a systematic variation on the HBT fit
parameters $\lambda$, $R_{o}^2$, $R_{s}^2$, and $R_{l}^2$, since
it discriminates against low-$q$ pairs which carry the correlation
signal. This is a consequence of the non-Gaussianess of the
correlation function. If it were a perfect Gaussian, this cut
would not change the extracted parameters from the Gaussian fit,
it would only reduce the statistics in certain bins and therefore
the only effect would be an increase in the statistical errors.

\begin{figure}
\includegraphics[width=0.40\textwidth]{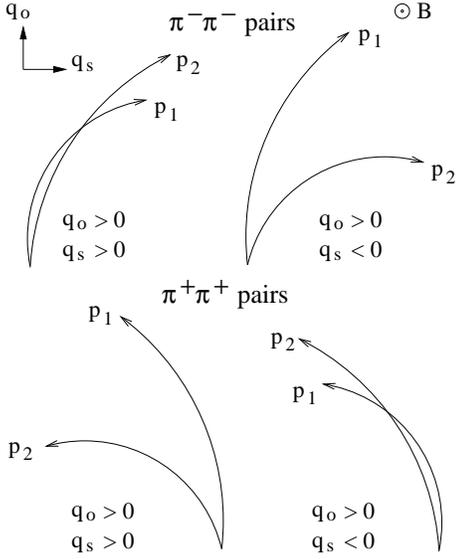}
\caption{For $\pi^-\pi^-$ ($\pi^+\pi^+$), merging occurs more
often between tracks with $|q_oq_s| = q_oq_s$ ($|q_oq_s| \neq
q_oq_s$) than with $|q_oq_s| \neq q_oq_s$ ($|q_oq_s| = q_oq_s$).
Note that exchanging the designations ``1" and ``2" does not
change the sign of $q_oq_s$.}\label{mergSk}
\end{figure}

In order to estimate this reduction we define a range in the
number of merged hits in which the lower limit is 0
(\textrm{i.e.}, no merging) and the higher limit is the value for
which we consider there is \textit{too much} merging. This value
is determined from the $0^{\textrm{th}}$ order Fourier
coefficients, $R_{os,0}^2$ which is expected to be 0,
Eq.~(\ref{FouCoef}). However, track merging introduces a deviation
of $R_{os}^2$ away from 0 caused by the preferential merging of
track pairs with correlated transverse momenta, $q_o$ and $q_s$ as
shown in Fig.~\ref{mergSk}. If we calculate the components of
$\vec{q}$ in the plane transverse to the beam as
$\vec{p}_{T,1}-\vec{p}_{T,2}$ where index 1 denotes the stiffer
track and define $\vec{\hat{\rho}}$ as the direction of the radius
of curvature of the stiffer track, then if
$\vec{q}\cdot\vec{\hat{\rho}}$ is positive there is more merging
on average. In the case of $\pi^-\pi^-$ pairs, there is a higher
degree of track merging when $|q_oq_s|$ = $q_oq_s$ than when
$|q_oq_s|$ $\neq$ $q_oq_s$ (top pairs). For $\pi^+\pi^+$ pairs the
conditions are opposite (bottom pairs).

When $R_{os}^2$ for $\pi^+$ or $\pi^-$ analysis clearly deviates
from 0, we consider that there is \textit{too much} merging and
use that value of the maximum fraction of merged hits as the upper
limit of the range. We calculate the change of each HBT radius in
this range and consider that to be the artificial reduction due to
the cut for that specific parameter. This reduction is included as
a systematic error in the final value. This is done for each
centrality and each $k_T$ bin.

\begin{figure}
\includegraphics[width=0.50\textwidth]{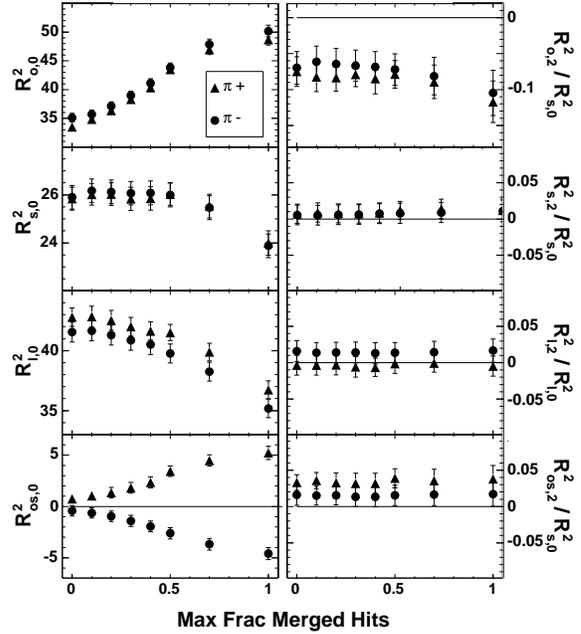}
\caption{Fourier coefficients as a function of the maximum
fraction of merged hits for the 5\% most central events and
$k_{T}$ between 150 and 250 MeV/c.}\label{SystErr_FC}
\end{figure}

As an example, Fig.~\ref{SystErr_FC} shows the $0^{\textrm{th}}$
order (left) and $2^{\textrm{nd}}$ divided by $0^{\textrm{th}}$
order (right) Fourier coefficients as a function of the maximum
fraction of merged hits allowed for the 5\% most central events
and 150  $<~k_{T}~<$ 250 MeV. From $R_{os,0}^2$, located in the
bottom left panel, we determined the upper limit of merged
fraction to be 0.2 and the corresponding variations in the HBT
radii to be 7\% for $R_{o}$, 5\% for $R_s$ and 10\% for $R_l$. The
systematic errors calculated according to this method are less or
equal than 10\% for all radii, in all centralities and $k_T$ bins.

\section{Pion HBT at $\sqrt{s{NN}}$ = 200 GeV}\label{results}

\subsection{How Gaussian is the measured correlation function?}\label{Gaussian}

\begin{figure}
\includegraphics[width=0.50\textwidth]{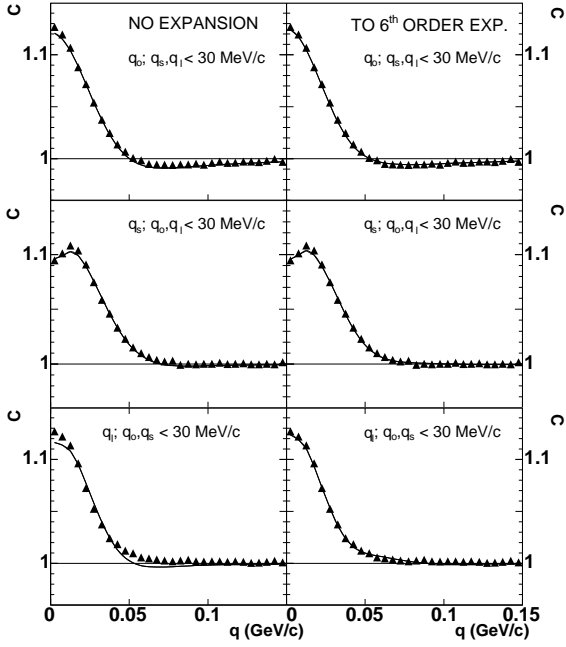}
\caption{Projections of the 3 dimensional correlation functions
and fits to Eq.~(\ref{BowlerFit}) (left) and with the Edgeworth
expansion to Eq.~(\ref{EdEx}) to $6^{\textrm{th}}$ order
(right).}\label{Projs_EdEx}
\end{figure}

\begin{figure}
\includegraphics[width=0.50\textwidth]{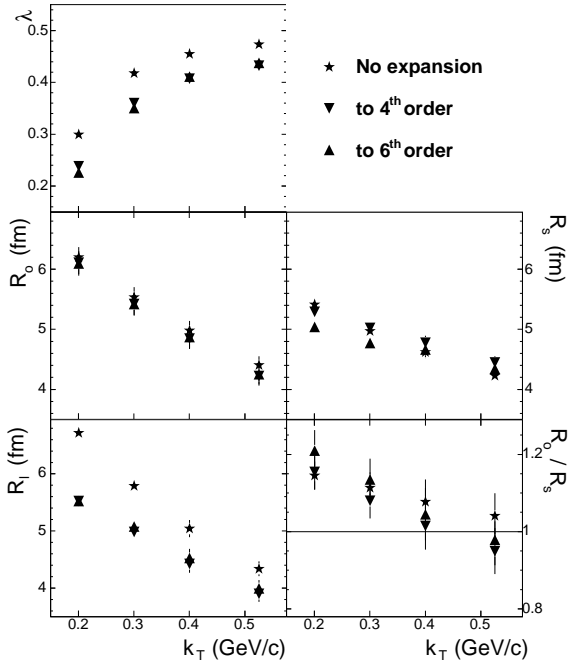}
\caption{HBT parameters for 0--5\% most central events for fits to
Eq.~(\ref{BowlerFit}) and to Eq.~(\ref{EdEx}) to $4^{\textrm{th}}$
and to $6^{\textrm{th}}$ orders. Error bars reflect only
statistical uncertainties.}\label{HBTpar_EdEx}
\end{figure}

Interferometric length scales are usually extracted from measured
correlation functions by fitting to a Gaussian functional form, as
discussed in~\ref{AnalysisMethod}. However, there is no reason to
expect the measured correlation function to be {\it completely}
Gaussian, and it is well-known that it seldom is.
Seemingly-natural questions such as ``how non-Gaussian is the
correlation function?'' ``how does the non-Gaussianness affect
extracted length scales?'' or ``what {\it is} the shape of the
correlation function?'' have, unfortunately, no unique,
assumption-free answers.

Often, non-Gaussian features are simply ignored in experimental
analyses.  Alternatively, the effect of non-Gaussianness is
estimated (e.g. by varying the range of $q$-values used in the
fit) and quoted as a systematic error on the HBT radii.
Occasionally, some alternative functional form (e.g. a sum of two
Gaussians, or an exponential plus a Gaussian) is chosen ad-hoc by
the experimenter based on the general ``appearance'' of the data.

One disadvantage of this last approach is the difficulty in
systematically comparing HBT results obtained with different
functional forms.  Since one might hope for evidence of ``new''
physics at RHIC, it is important to place RHIC HBT results into
the context of previously-established systematics.  Thus this
paper mainly focusses on Gaussian HBT radius systematics. However,
to address non-Gaussian issues, here we move beyond ad-hoc methods
and adopt as a standard the {\it Edgeworth expansion} proposed by
Cs\"{o}rg\H{o} and collaborators~\cite{CSORG99,CSORG00,CSORG03}.
Using the \textit{Bowler-Sinyukov} Coulomb treatment, the
correlation functions are fitted to
\begin{eqnarray}\label{EdEx}
\lefteqn{C(q_o,q_s,q_l) = (1 - \lambda) + \lambda K_{\textrm{coul}}(q_{\textrm{inv}})}\nonumber \\
& & + \lambda K_{\textrm{coul}}(q_{\textrm{inv}})\cdot e^{-q_o^2R_o^2-q_s^2R_s^2-q_l^2R_l^2}\times\nonumber \\
& & \Big[1 + \sum_{n=4,
\textrm{n~even}}^{\infty}\frac{\kappa_{o,n}}{n!(\sqrt{2})^n}H_n(q_oR_o)\Big]\times\nonumber \\
& & \Big[1 + \sum_{n=4,
\textrm{n~even}}^{\infty}\frac{\kappa_{s,n}}{n!(\sqrt{2})^n}H_n(q_sR_s)\Big]\times\nonumber \\
& & \Big[1 + \sum_{n=4,
\textrm{n~even}}^{\infty}\frac{\kappa_{l,n}}{n!(\sqrt{2})^n}H_n(q_lR_l)\Big],
\end{eqnarray}
where $\kappa_{i,n}$ ($i =o,s,l$) are fit parameters and
$H_n(q_iR_i)$ are the Hermite polynomials of order \textit{n}:
\begin{equation}
H_n (x) = (-1)^n e^{x^2}\frac{d^n}{dx^n}~e^{-x^2}.
\end{equation}
Only Hermite polynomials of even order are included in the
expansion because the correlation function for identical particles
must be invariant under
$(q_o,q_s,q_l)~\rightarrow~(-q_o,-q_s,-q_l)$.

At midrapidity and integrated over azimuthal angle, the quantum
interference term is factorizable into the $q_o$, $q_s$, and $q_l$
variables. Therefore, it may be uniquely decomposed in terms of
{\it any} complete set of basis functions of these variables.
Given a sufficient number of terms, any basis set will do. Thus,
the potential advantage of the Edgeworth decomposition
(Eq.~\ref{EdEx}) is {\it not} that it is any more
``model-independent''~\cite{CSORG00} than, say, a Tschebyscheff
decomposition, but that a functional expansion about a Gaussian
shape might most economically describe the data. Since they are
approximately Gaussian, one hopes to capture the shape of measured
correlation functions with only a few low-order terms.

\begin{table*}
\begin{tabular}{l|c|c|c|c}
 \hline
$k_T$ (MeV/c)& 150--250 & 250--350 & 350--450 & 450--600 \\
\hline %
$\lambda$ & 0.30 $\pm$ 0.01 & 0.42 $\pm$ 0.01 & 0.45 $\pm$ 0.01 & 0.47 $\pm$ 0.01\\
$\lambda$ ($4^{\textrm{th}}$ ord.) & 0.24 $\pm$ 0.01 & 0.36 $\pm$ 0.01 & 0.41 $\pm$ 0.01 & 0.43 $\pm$ 0.01\\
$\lambda$ ($6^{\textrm{th}}$ ord.) & 0.23 $\pm$ 0.01 & 0.35 $\pm$ 0.01 & 0.41 $\pm$ 0.01 & 0.44 $\pm$ 0.01\\
\hline \hline
$R_o$ & 6.16 $\pm$ 0.01 & 5.51 $\pm$ 0.01 & 4.88 $\pm$ 0.02 & 4.32 $\pm$ 0.02\\
\hline
$R_o$ ($4^{\textrm{th}}$ ord.) & 6.07 $\pm$ 0.04 & 5.40 $\pm$ 0.03 & 4.75 $\pm$ 0.03 & 4.14 $\pm$ 0.04\\
$\kappa_{o,4}$ & 0.37 $\pm$ 0.05 & 0.36 $\pm$ 0.04 & 0.33 $\pm$ 0.05 & 0.40 $\pm$ 0.06\\
\hline%
$R_o$ ($6^{\textrm{th}}$ ord.) & 6.05 $\pm$ 0.05 & 5.40 $\pm$ 0.04 & 4.78 $\pm$ 0.04 & 4.17 $\pm$ 0.04\\
$\kappa_{o,4}$ & 0.53 $\pm$ 0.11 & 0.45 $\pm$ 0.10 & 0.20 $\pm$ 0.11 & 0.22 $\pm$ 0.13\\
$\kappa_{o,6}$ & 0.83 $\pm$ 0.39 & 0.53 $\pm$ 0.38 & 0.63 $\pm$ 0.44 & --0.84 $\pm$ 0.53\\
\hline\hline
$R_s$ & 5.39 $\pm$ 0.01 & 4.93 $\pm$ 0.01 & 4.53 $\pm$ 0.01 & 4.14 $\pm$ 0.02\\
\hline %
$R_s$ ($4^{\textrm{th}}$ ord.) & 5.27 $\pm$ 0.03 & 4.98 $\pm$ 0.03 & 4.68 $\pm$ 0.03 & 4.36 $\pm$ 0.03\\
$\kappa_{s,4}$& 0.22 $\pm$ 0.04 & --0.03 $\pm$ 0.04 & --0.27 $\pm$ 0.04 & --0.50 $\pm$ 0.05\\
\hline %
$R_s$ ($6^{\textrm{th}}$ ord.) & 5.01 $\pm$ 0.05 & 4.74 $\pm$ 0.04 & 4.57 $\pm$ 0.04 & 4.26 $\pm$ 0.04\\
$\kappa_{s,4}$ & 0.99 $\pm$ 0.10 & 0.79 $\pm$ 0.10 & 0.16 $\pm$ 0.11 & --0.07 $\pm$ 0.13\\
$\kappa_{s,6}$ & 3.07 $\pm$ 0.35 & 3.21 $\pm$ 0.37 & 1.71 $\pm$ 0.44 & 1.80 $\pm$ 0.51\\
\hline\hline
$R_l$ & 6.64 $\pm$ 0.02 & 5.72 $\pm$ 0.02 & 4.94 $\pm$ 0.02 & 4.25 $\pm$ 0.02\\
\hline %
$R_l$ ($4^{\textrm{th}}$ ord.) & 5.47 $\pm$ 0.04 & 4.92 $\pm$ 0.03 & 4.33 $\pm$ 0.04 & 3.82 $\pm$ 0.04\\
$\kappa_{l,4}$ & 1.60 $\pm$ 0.06 & 1.25 $\pm$ 0.05 & 1.04 $\pm$ 0.06 & 0.78 $\pm$ 0.06\\
\hline%
$R_l$ ($6^{\textrm{th}}$ ord.) & 5.01 $\pm$ 0.05 & 5.01 $\pm$ 0.04 & 4.43 $\pm$ 0.04 & 3.91 $\pm$ 0.04\\
$\kappa_{l,4}$ & 1.32 $\pm$ 0.07 & 0.70 $\pm$ 0.07 & 0.54 $\pm$ 0.09 & 0.32 $\pm$ 0.11\\
$\kappa_{l,6}$ & --1.76 $\pm$ 0.29 & --2.82 $\pm$ 0.29 & --2.41 $\pm$ 0.35 & --2.12 $\pm$ 0.43\\
\hline
\end{tabular}
\caption{HBT parameters and $\kappa$ parameters for fits of the
correlations functions without and up to $4^{\textrm{th}}$ and
$6^{\textrm{th}}$ order of the Edgeworth expansion for the 5\%
most central events. The extracted radii are also shown in
Fig.~\ref{HBTpar_EdEx}.} \label{EdExTable}
\end{table*}

\begin{table}
\begin{tabular}{c|c|c|c}
\hline
$k_T$ (MeV/c) & No exp. & To $4th$ order & To $6th$ order\\
\hline
150--250 & 1.23 & 1.09 & 1.09\\
\hline
250--350 & 1.22 & 1.05 & 1.04\\
\hline
350--450 & 1.20 & 1.02 & 1.02\\
\hline
450--600 & 1.17 & 1.01 & 1.01\\
\hline
\end{tabular}
\caption{$\chi^2$/dof for fits of the correlations functions
without and up to $4^{\textrm{th}}$ and $6^{\textrm{th}}$ order of
the Edgeworth expansion for the 5\% most central events.}
\label{EdExChi2}
\end{table}

We fit our correlation functions to the form given by
Eq.~(\ref{EdEx}) for two different cases, up to $n=4$ and up to
$n=6$ of the Hermite polynomials, and compare with fits to Eq.~
(\ref{BowlerFit}) (without expansion). In Fig.~\ref{Projs_EdEx} we
show the fits to projections of the correlation function for the
0--5\% most central events and $k_T$ between 150 and 250 MeV/c,
with no expansion in the left column and with expansion up to
$6^{\textrm{th}}$ order in the right column. We observe a small
improvement in the fit when we include the expansion. In
Fig.~\ref{HBTpar_EdEx} the extracted HBT parameters as a function
of $k_T$ for the 0--5\% most central events for the fits without
expansion, with expansion up to $4^{\textrm{th}}$ order and with
expansion up to $6^{\textrm{th}}$ order are shown. In Table
\ref{EdExTable} are the corresponding values for the $\kappa$
parameters. When comparing the extracted parameters including the
expansion to $6^{\textrm{th}}$ order to those extracted without
the expansion, we observe that $R_o$ decreases by $\sim$~2\% for
all $k_T$ bins, $R_s$ changes between $\sim$~--7\% for the lowest
$k_T$ bin [150,250] MeV/c and $\sim$~+3\% for the highest bin
[450,600] MeV/c, and $R_l$ decreases between $\sim$~18\% and
$\sim$~8\% for the lowest and highest $k_T$ bins respectively. In
Table \ref{EdExChi2} are the corresponding $\chi^2/dof$ for those
same fits. $\chi^2$/dof slightly improves when including the
expansion up to $4^{th}$ order and does not change with the
expansion to $6^{th}$ order. Similar trends are observed at all
centralities.

We do not consider the change in HBT radii when including an
Edgeworth expansion to represent a systematic uncertainty when
comparing to Gaussian radii traditionally discussed in the
literature. Rather, the differences reflects a deviation from the
Gaussian shape traditionally assumed. Furthermore, the expansion
provides a more detailed, yet still compact, characterization of
the measured correlation function. Further theoretical development
of the formalism, outside the scope of this paper, is required to
determine whether the expansion parameters convey important
physical information beyond that carried by Gaussian radius
parameters.

\subsection{$m_{T}$ dependence of the HBT parameters for most central
collisions}\label{C1results}

The HBT radius parameters measure the sizes of the homogeneity
regions (regions emitting particles of a given momentum)
\cite{MAKHL88}. Hence, for an expanding source, depending on the
momenta of the pairs of particles entering the correlation
function, different parts of the source are measured. The size of
these regions are controlled by the velocity gradients and
temperature \cite{WIEDE98,TOMAS00,WIEDE96}. Therefore the
dependence of the transverse radii on transverse mass $m_{T}$
contains dynamical information of the particle emitting source
\cite{PRATT84,MAKHL88}.

\begin{figure}
\includegraphics[width=0.50\textwidth]{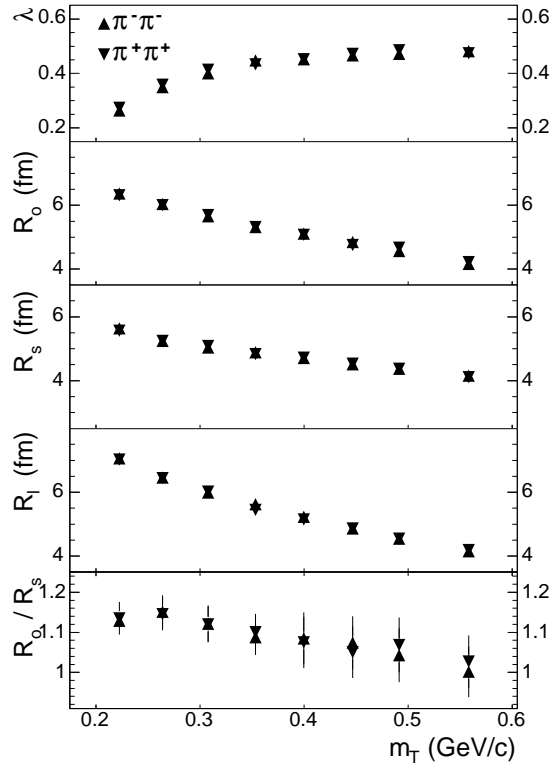}
\caption{HBT parameters for 0--5\% most central events for
$\pi^+\pi^+$ and $\pi^-\pi^-$ correlation functions. Error bars
include statistical and systematic uncertainties.}
\label{HBTpar_vsmT_C1}
\end{figure}

Figure \ref{HBTpar_vsmT_C1} shows the HBT parameters $\lambda$,
$R_{o}$, $R_{s}$, $R_{l}$ and the ratio $R_{o}/R_{s}$ for the
0--5\% most central events as a function of $m_{T}$ for
$\pi^+\pi^+$ and $\pi^-\pi^-$ correlation functions. We observe
excellent agreement between the parameters extracted from the
positively and negatively charged pion analyses. The $\lambda$
parameter increases with $m_{T}$. This is consistent with studies
at lower energies \cite{ADLER01, BEARD98,LISA00,AGGAR00}, in which
the increase was attributed to decreased contributions of pions
from long-lived resonances at higher $p_T$. The three HBT radii
rapidly decrease as a function of $m_{T}$; the decrease of the
transverse radii ($R_o$ and $R_s$) with $m_T$ is usually
attributed to the radial flow \cite{WIEDE98,TOMAS00,WIEDE96}; the
strong decrease in $R_l$ might be produced by the longitudinal
flow \cite{WIEDE98,MAKHL88,WIEDE96,AKKEL95,HERRM95}. $R_o$ falls
steeper than $R_s$ with $m_T$ which is consistent with $R_o$ being
more affected by radial flow \cite{RETIE03}. In contrast to many
model predictions \cite{RISCH96,HEINZ02}, $R_{o}/R_{s}~\sim~1$
which indicates short emission duration in a blast wave
parametrization \cite{RETIE03} as will be discussed in next
section.

\begin{figure}
\includegraphics[width=0.50\textwidth]{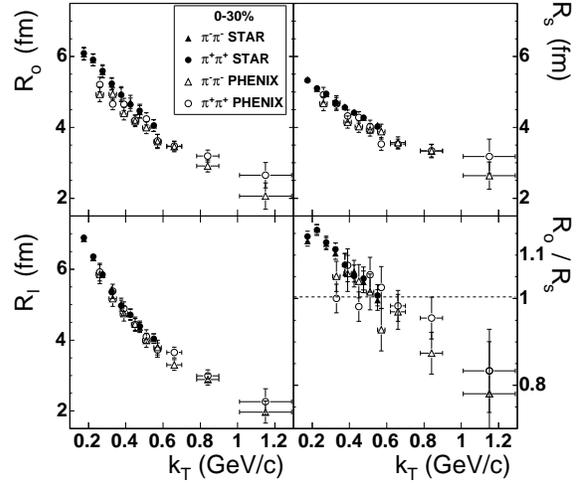}
\caption{HBT parameters from STAR and PHENIX at the same beam
energy for the 0--30\% most central events. Error bars include
statistical and systematic uncertainties.}\label{STARvsPHENIX}
\end{figure}

Figure \ref{STARvsPHENIX} compares our extracted HBT radius
parameters from $\pi^+\pi^+$ and $\pi^-\pi^-$ correlation
functions for the 0--30\% most central events with those obtained
by the PHENIX collaboration \cite{ADLER03} at the same beam energy
and centrality. The same fitting procedure has been used in both
analysis. In general, very good agreement is observed in the three
radii, although small discrepancies are seen in $R_o$ at small
$k_T$.

\begin{figure}
\includegraphics[width=0.50\textwidth]{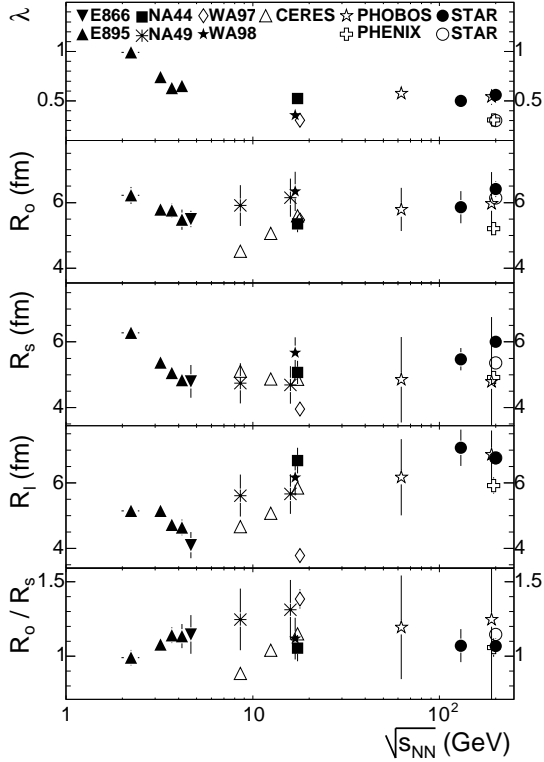}
\caption{Energy dependence of $\pi^-$ HBT parameters for central
Au+Au, Pb+Pb, and Pb+Au collisions at midrapidity and $\langle
k_T\rangle~\approx~0.2$ GeV/c
\cite{ADLER01,ADAMO03,BEARD98,LISA00,AGGAR00,ADLER03,SOLTZ99,BLUME02,ANTIN01,BACK04}.
Open symbols indicate that fitting was done according to the
\textit{Bowler-Sinyukov} procedure (or a similar one in the case
of the results from Phobos). Error bars on NA44, NA49, CERES,
PHENIX, Phobos and STAR results include systematic uncertainties;
error bars on other results are only statistical.}
\label{ExcitFunction}
\end{figure}

Figure \ref{ExcitFunction} shows the HBT parameters vs. collision
energy for midrapidity, low $p_T$ $\pi^-\pi^-$ from central Au+Au,
Pb+Pb or Pb+Au collisions. In order to compare with our previous
results at $\sqrt{s_{NN}} = 130$ GeV, we applied similar cuts in
our analysis as those described in \cite{ADLER01} and fit our
correlation function according to the \textit{standard} procedure
described in section \ref{CC} to extract the HBT parameters at
$\sqrt{s_{NN}} = 200$ GeV, closed circles at that energy in
Fig.~\ref{ExcitFunction}. We observe an increase of $\sim$~10\% in
the transverse radii $R_o$ and $R_s$. In the case of $R_s$, this
increase could be attributed to a larger freeze-out volume for a
larger pion multiplicity. $R_l$ is consistent with our result at
lower energy. The predicted increase by hydrodynamic models in the
ratio $R_o/R_s$ as a probe of the formation of QGP is not observed
at $\sqrt{s_{NN}} = 200$ GeV. More discussion on the lack of
energy dependence of the HBT radii and its possible relation with
the constant mean free path can be found in reference
\cite{ADAMO03b}.

We have also included in Fig.~\ref{ExcitFunction} the values for
the HBT parameters at $\sqrt{s_{NN}} = 200$ GeV extracted when
applying the cuts discussed in section \ref{ExpSetup} and fitting
the correlation function according to the \textit{Bowler-Sinyukov}
procedure (section \ref{CC}), open circles in the figure. This
procedure is also used by the CERES collaboration. The smaller
$\lambda$, $R_o$, and $R_s$ can be explained by the different cuts
as already discussed in section \ref{CC}. The larger value for
$R_o/R_s$ is due to the improved procedure of taking Coulomb
interaction into account in the \textit{Bowler-Sinyukov}
procedure, section \ref{CC}.

\subsection{Centrality dependence of the $m_{T}$ dependence}\label{CentMtDepen}

\begin{figure}
\includegraphics[width=0.45\textwidth]{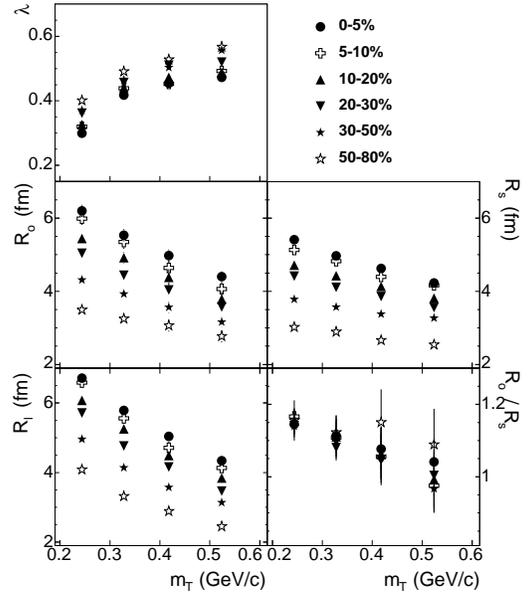}
\caption{HBT parameters vs. $m_T$ for 6 different centralities.
Error bars include statistical and systematic
uncertainties.}\label{HBTpar_vsmT_allCents}
\end{figure}

We observe excellent agreement between the results for positively
and negatively charged pion correlation functions for the most
central collisions shown in the previous section. Therefore, we
add the numerators and denominators of the correlation functions
for positive and negative pions in order to improve statistics;
all the results shown in the rest of this section correspond to
these added correlation functions. The centrality dependence of
the source parameters is presented in
Fig.~\ref{HBTpar_vsmT_allCents} where the HBT parameters are shown
as a function of $m_{T}$ for 6 different centralities. The
$\lambda$ parameter slightly increases with decreasing centrality.
The three radii increase with increasing centrality and $R_{l}$
varies similar to $R_{o}$ and $R_{s}$. For $R_{o}$ and $R_{s}$
this increase may be attributed to the initial geometrical overlap
of the two nuclei. $R_{o}/R_{s}~\sim~1$, for all centralities.

\subsection{Azimuthally-sensitive HBT}\label{asHBT}

\begin{figure}
\includegraphics[width=0.50\textwidth]{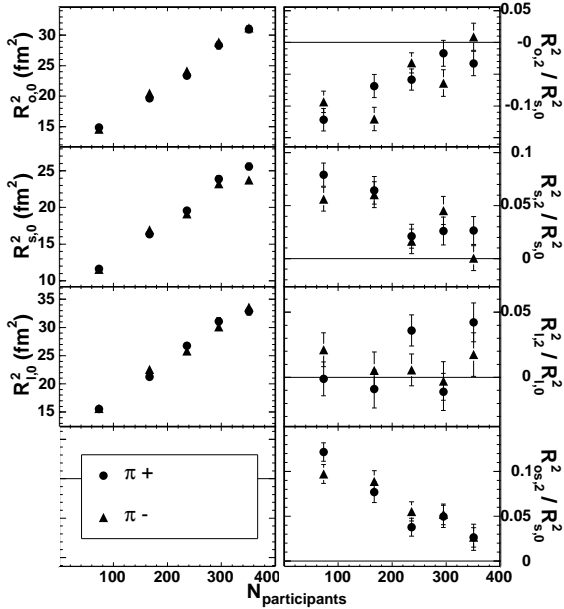}
\caption{Fourier coefficients of azimuthal oscillations of HBT
radii vs.~number of participating nucleons, for $\pi^+$ and
$\pi^-$ pairs separately ($0.25 < k_T < 0.35$ GeV/c).  Left
panels: means ($0^{\rm{th}}$-order FC) of oscillations; right
panels: relative amplitudes (see text for details).  Larger
participant numbers correspond to more central
collisions.}\label{comparePions}
\end{figure}

The results presented in \cite{ADAMS03c} were for $\pi^+$ and
$\pi^-$ correlation functions combined before fitting.
Figure~\ref{comparePions} shows the consistency of the Fourier
coefficients obtained with Eq.~(\ref{FouCoef}) as a function of
number of participants.

\begin{figure}
\includegraphics[width=0.50\textwidth]{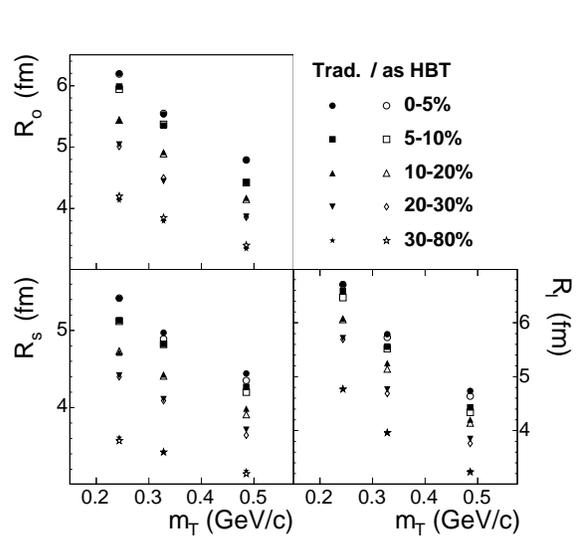}
\caption{Comparison between the HBT radii obtained from and
azimuthally integrated (\textit{traditional}) HBT analysis and the
$0^{\textrm{th}}$-order Fourier coefficients from an
azimuthally-sensitive HBT analysis. Error bars include only
statistical uncertainties.}\label{StvsAs}
\end{figure}

In section \ref{AnalysisMethod} we noted that the
$0^{\textrm{th}}$ order Fourier coefficients correspond to the
extracted HBT radii in an azimuthally integrated analysis. This is
confirmed in Fig.~\ref{StvsAs} that shows the excellent agreement
between them. The azimuthally integrated (\textit{traditional})
HBT radii (closed symbols) agree within 1/10 fm with the
$0^{\textrm{th}}$-order Fourier coefficients (open symbols) from
the azimuthally-sensitive HBT analysis.

\section{Discussion of results}\label{Discussion}

\begin{figure}
\includegraphics[width=0.50\textwidth]{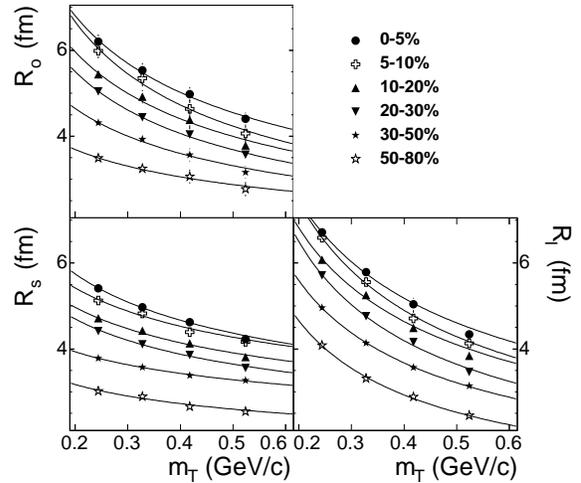}
\caption{HBT radius parameters for 6 different centralities. The
lines indicate power-law fits ($R_{i}(m_{T})~=~R_i'\cdot
(m_{T}/m_{\pi})^{-\alpha_i}$) to each parameter for each
centrality.}\label{HBTpar_vsmT_powerLawFits}
\end{figure}

\begin{figure}
\includegraphics[width=0.50\textwidth]{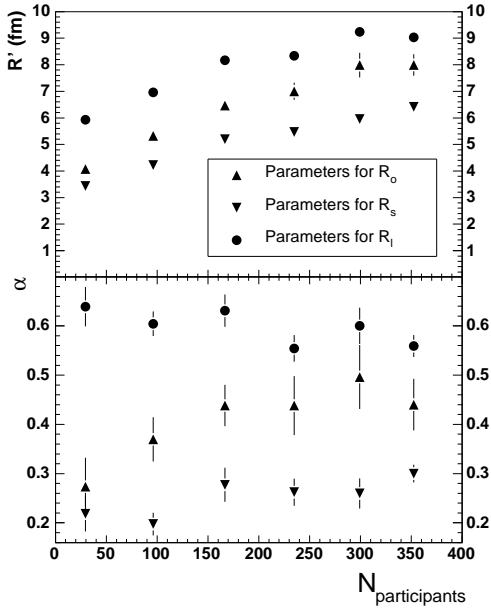}
\caption{Extracted parameters $R'$ in the top panel, $\alpha$ in
the bottom from the power-law fits to the HBT radius parameters
(lines in Fig.~\ref{HBTpar_vsmT_powerLawFits}).}
\label{R0_alpha_vsNpart}
\end{figure}

As already mentioned in section \ref{C1results}, the dependence of
the HBT radii on the transverse mass contains dynamical
information about the particle emitting source. In order to
extract this information, we fit the $m_T$ dependence of the HBT
radii for each centrality from Fig.~\ref{HBTpar_vsmT_allCents}
using a simple power-law fit: $R_{i}(m_{T})~=~R_i'\cdot
(m_{T}/m_{\pi})^{-\alpha_i}$ (solid lines in
Fig.~\ref{HBTpar_vsmT_powerLawFits}). Figure
\ref{R0_alpha_vsNpart} shows the extracted fit parameters for the
three HBT radii, $R'$ in the top panel and $\alpha$ in the lower
panel, as a function of the number of participants, where
$N_{\textrm{participants}}$ has been calculated from a Glauber
model described in \cite{ADAMS03b}. $N_{\textrm{participants}}$
increases with the centrality of the collision. $R'$ decreases
with decreasing number of participants, which is consistent with
the decreasing initial source size. $\alpha$ is approximately
constant for $R_l$ which would indicate that the longitudinal flow
is similar for all centralities. However, for the transverse radii
$R_o$ and $R_s$, $\alpha$ seems to decrease for the most
peripheral collisions which could be an indication of a small
reduction of transverse flow and/or an increase of temperature for
those most peripheral collisions. This is consistent with the
values for flow and temperature extracted from blast wave fits to
pion, kaon, and proton transverse momentum spectra \cite{ADAMS03},
as well as to HBT as will be discussed in next section. The drop
of $\alpha$ with decreasing number of participants is faster in
$R_o$ than in $R_s$ which could again indicate that $R_o$ might be
more affected by radial flow \cite{RETIE03}.

\subsection{Blast wave parametrization}\label{BW}

\begin{table*}
\begin{tabular}{|c|c|c|c|c|c|c|}
  \hline
  Centrality (\%) & $T$ (MeV) & $\rho_0$ & $R$ (fm) & $\tau$ (fm/c) & $\Delta\tau$ (fm/c) & $\chi^2$/dof\\
  \hline
  0--5   &  97 $\pm$ 2 & 1.03 $\pm$ 0.01 & 13.3 $\pm$ 0.2 & 9.0 $\pm$ 0.3 & 2.83 $\pm$ 0.19 & 3.13/9\\
  5--10  &  98 $\pm$ 2 & 1.00 $\pm$ 0.01 & 12.6 $\pm$ 0.2 & 8.7 $\pm$ 0.2 & 2.45 $\pm$ 0.17 & 2.71/9\\
  10--20 &  98 $\pm$ 3 & 0.98 $\pm$ 0.01 & 11.5 $\pm$ 0.2 & 8.1 $\pm$ 0.2 & 2.35 $\pm$ 0.16 & 2.61/9\\
  20--30 & 100 $\pm$ 2 & 0.94 $\pm$ 0.01 & 10.5 $\pm$ 0.1 & 7.2 $\pm$ 0.1 & 2.10 $\pm$ 0.09 & 0.99/9\\
  30--50 & 108 $\pm$ 2 & 0.86 $\pm$ 0.01 &  8.8 $\pm$ 0.1 & 5.9 $\pm$ 0.1 & 1.74 $\pm$ 0.12 & 2.13/9\\
  50--80 & 113 $\pm$ 2 & 0.74 $\pm$ 0.01 &  6.5 $\pm$ 0.1 & 4.0 $\pm$ 0.2 & 1.73 $\pm$ 0.10 & 1.12/9\\
  \hline
\end{tabular}
\caption{Extracted parameters from a blast wave fit to azimuthally
integrated pion HBT radii, with $T$ and $\rho_0$ fixed from fits
to pion, kaon, and proton transverse momentum spectra and
$v_2$.}\label{BWtable1}
\begin{tabular}{|c|c|c|c|c|c|c|c|c|}
  \hline
  Cent. (\%) & $T$ (MeV) & $\rho_0$ & $\rho_2$ & $R_x$ (fm) &$R_y$ (fm) & $\tau$ (fm/c) & $\Delta\tau$ (fm/c) & $\chi^2$/dof\\
  \hline
  0--5   &  97 $\pm$ 2 & 1.03 $\pm$ 0.01 & 0.03 $\pm$ 0.002 & 12.9 $\pm$ 0.1 & 13.4 $\pm$ 0.1 & 8.9 $\pm$ 0.2 & 3.16 $\pm$ 0.11 & 106.8/63\\
  5--10  &  98 $\pm$ 2 & 1.00 $\pm$ 0.01 & 0.04 $\pm$ 0.002 & 12.1 $\pm$ 0.1 & 12.9 $\pm$ 0.1 & 8.2 $\pm$ 0.2 & 2.73 $\pm$ 0.12 & 103.2/63\\
  10--20 &  98 $\pm$ 3 & 0.98 $\pm$ 0.01 & 0.05 $\pm$ 0.002 & 10.9 $\pm$ 0.1 & 11.9 $\pm$ 0.1 & 7.8 $\pm$ 0.2 & 2.59 $\pm$ 0.10 & 131.06/63\\
  20--30 & 100 $\pm$ 1 & 0.94 $\pm$ 0.01 & 0.07 $\pm$ 0.002 &  9.7 $\pm$ 0.1 & 11.0 $\pm$ 0.1 & 6.9 $\pm$ 0.1 & 2.29 $\pm$ 0.10 & 87.2/63\\
  30--80 & 112 $\pm$ 2 & 0.82 $\pm$ 0.05 & 0.10 $\pm$ 0.005 &  7.4 $\pm$ 0.1 &  8.7 $\pm$ 0.1 & 5.1 $\pm$ 0.2 & 1.94 $\pm$ 0.14 & 189.4/63\\
  \hline
\end{tabular}
\caption{Extracted parameters from a blast wave fit to azimuthally
sensitive pion HBT radii, with $T$ and $\rho_0$ fixed from fits to
pion, kaon, and proton transverse momentum spectra and $v_2$.}
\label{BWtable2}
\end{table*}

Hydrodynamic calculations that successfully reproduce transverse
momentum spectra and elliptic flow, fail to reproduce the HBT
parameters \cite{HEINZ02}. In most cases, these calculations
underestimate $R_s$ and overestimate $R_o$ and $R_l$. Since $R_s$
only probes the spatial extent of the source while $R_o$ and $R_l$
are also sensitive to the system lifetime and the duration of the
particle emission \cite{PRATT84}, they may be underestimating the
system size and overestimating its evolution time and emission
duration. We fit our data with a blast wave parametrization
designed to describe the kinetic freeze-out configuration. In this
section we will discuss the extracted parameters and their
physical implications.

This blast wave parametrization \cite{RETIE03} assumes that the
system is contained within an infinitely long cylinder along the
beam line and requires longitudinal boost invariant flow. It will
be shown that this latter assumption is not necessarily correct.
It also assumes uniform particle density. The single set of free
parameters in this parametrization is: the kinetic freeze-out
temperature ($T$); the maximum flow rapidity ($\rho =
\tilde{r}(\rho_0+\rho_a \cos(2\phi))$, for an azimuthally
integrated analysis $\rho_a = 0$); the radii ($R$ for the
azimuthally integrated analysis and $R_x, R_y$ for the azimuthally
sensitive analysis) of the cylindrical system; the system
longitudinal proper time ($\tau = \sqrt{t^2 - z^2}$); and the
emission duration ($\Delta\tau$).

We use this parametrization to fit the azimuthally integrated pion
HBT radii, as well as the azimuthally sensitive pion HBT radii. In
both fits, $T$ and $\rho_0$ are fixed to those extracted from a
blast wave fit to pion, kaon and proton transverse momentum
spectra \cite{ADAMS03} and $v_2$ \cite{ADAMS04}. By doing this,
the azimuthally integrated and azimuthally sensitive radii are
fitted with the same temperature and $\rho_0$, and the edge source
radii can be compared directly. Also, a 5\% error was added to all
HBT radii before the fit in order to reflect the blast wave
systematic errors described in \cite{RETIE03}.  In the fit, the
transverse flow rapidity linearly increases from zero at the
center to a maximum value at the edge of the system.The best fit
parameters are summarized in Table \ref{BWtable1}, for the
azimuthally integrated analysis, and Table \ref{BWtable2}, for the
azimuthally sensitive analysis.

Most of the parameters, as well as their evolution with
centrality, agree with similar studies. Temperature decreases with
increasing centrality and the average transverse flow velocity
($\langle\beta_T\rangle = \int
\textrm{arctanh}(\rho_0\frac{r}{R})rdr/\int rdr$) increases with
increasing centrality. Both results are consistent with those
extracted from fits to spectra only \cite{ADAMS03} and reflect
increased rescattering, expansion and system evolution time with
increasing centrality.

\begin{figure}
\includegraphics[width=0.50\textwidth]{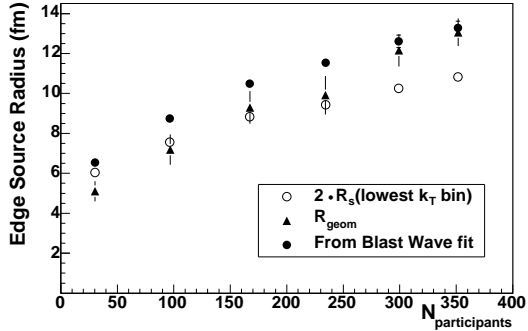}
\caption{Extracted freeze-out source radius extracted from a blast
wave fit; source radius $R_{\textrm{geom}}$ from fits to $R_s$
(lines in Fig.~\ref{RsFit}); and $2\cdot R_s$ for the lowest $k_T$
bin as a function of number of participants. $R$ from blast wave
contains only uncertainties from fit; $R_{\textrm{geom}}$ error
bars contain systematic uncertainties from the input parameters;
$2\cdot R_s$ contains the statistical and systematic uncertainties
from $R_s$.}\label{Rgeom}
\end{figure}

\begin{figure}
\includegraphics[width=0.50\textwidth]{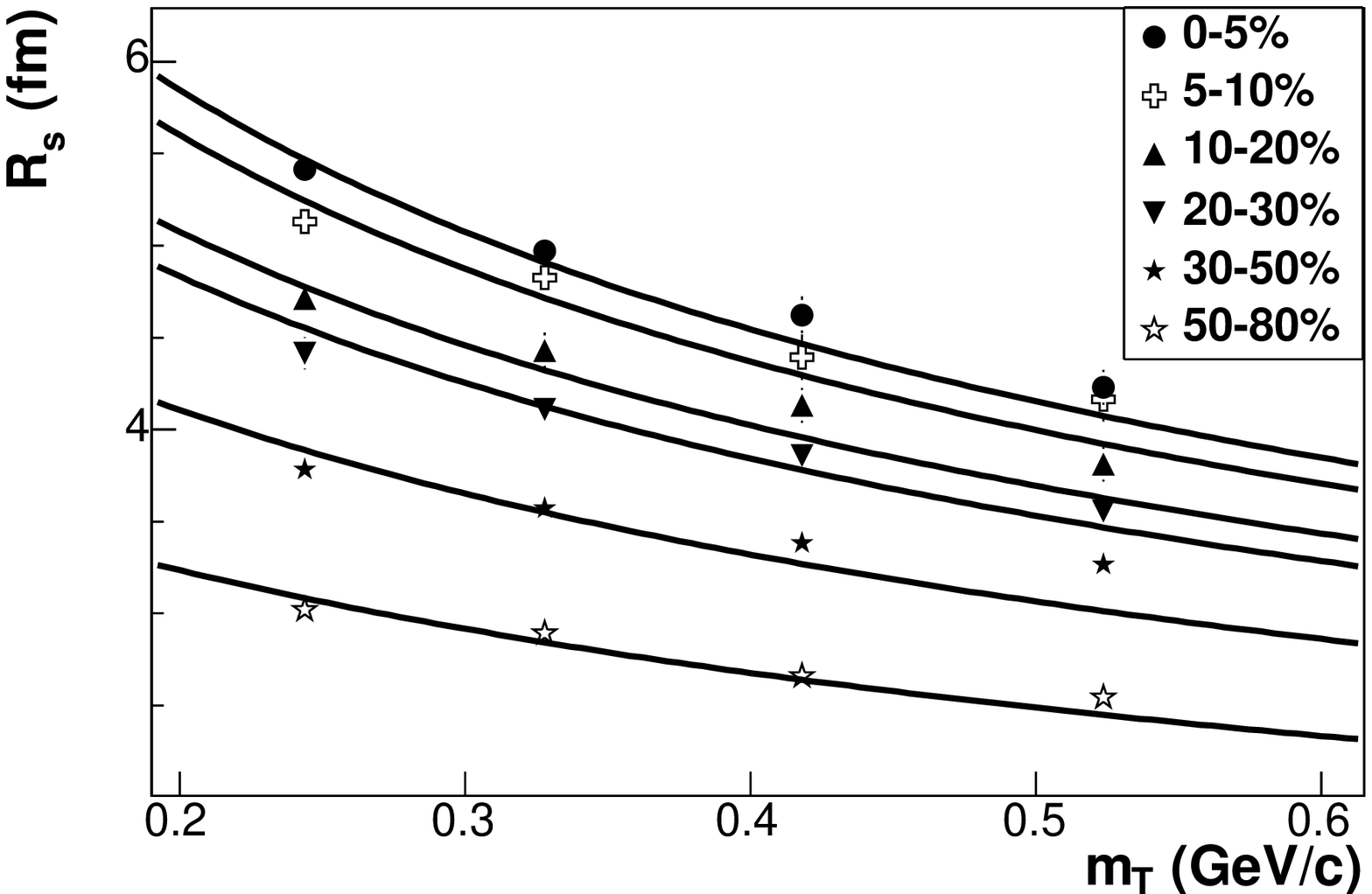}
\caption{HBT parameter $R_s$. Lines represent the fits
$R_{s}(m_{T}) = \sqrt{R_{\textrm{geom}}^{2}/\Big[1 +
\rho_{0}^{2}(\frac{1}{2} + \frac{m_{T}}{T})\Big]}$.}\label{RsFit}
\end{figure}

Figure \ref{Rgeom} shows the $R$ parameter extracted from the
blast wave fit as they are in Table \ref{BWtable1}. Also shown in
that plot is $R_{\textrm{geom}}$ calculated assuming a transverse
expanding, longitudinally boost-invariant source, and a Gaussian
transverse density profile, by fitting the $m_{T}$ dependence of
$R_{s}$ to \cite{WIEDE96}:
\begin{equation}
R_{s}(m_{T}) = \sqrt{\frac{R_{\textrm{geom}}^{2}}{1 +
\rho_{0}^{2}(\frac{1}{2} + \frac{m_{T}}{T})}},
\end{equation}
where $T$ is the freeze-out temperature and $\rho_{0}$ is the
surface transverse rapidity. Figure \ref{RsFit} shows such fits to
$R_s$ for each centrality with $T$ and $\rho_0$ extracted from
blast wave fits to pion, kaon, and proton transverse momentum
spectra ($T$ = 90 MeV, $\rho_{0}$ = 1.20 for the most central
collisions and $T$ = 120 MeV, $\rho_{0}$ = 0.82 for the most
peripheral bins) \cite{ADAMS03}. Fig.~\ref{Rgeom} shows good
agreement between these two extracted radii that increase from
$\sim$~5 fm for the most peripheral collisions to $\sim$~13 fm for
the most central collisions following the growth of the system
initial size. The differences may be explained by the poor quality
of the fits to $R_s$ as seen in Fig.~\ref{RsFit}.

As already mentioned, $R_s$ carries only spatial information about
the source \cite{WIEDE98,WIEDE99}. In the special case of
vanishing space-momentum correlations (no transverse flow or $T
\rightarrow \infty$), the source spatial distribution may be
modelled by a uniformly filled disk of radius $R$, which in this
case is exactly two times $R_s$, the RMS of the distribution along
a specific direction. In Fig.~\ref{Rgeom} we have included $2\cdot
R_s$ for our lowest $k_T$ bin, $k_T$ = [150,250] MeV/c, in order
to compare it with the extracted source radii. We observe the
effect of space-momentum correlations that reduce the size of the
regions of homogeneity in the results from the most central
collisions for which $2\cdot R_s$ is smaller than the extracted
radii from the fit.

\begin{table}
\begin{tabular}{|c|c|c|}
  \hline
  Centrality & $R_{x,\textrm{initial}}$ (fm) & $R_{y,\textrm{initial}}$ (fm) \\
  \hline
  0--5\% & 5.70 $\pm$ 0.01 & 5.86 $\pm$ 0.01 \\
  5--10\% & 5.28 $\pm$ 0.01 & 5.72 $\pm$ 0.01 \\
  10--20\% & 4.74 $\pm$ 0.01 & 5.50 $\pm$ 0.01 \\
  20--30\% & 4.14 $\pm$ 0.01 & 5.12 $\pm$ 0.01 \\
  30--50\% & 3.58 $\pm$ 0.01 & 4.70 $\pm$ 0.01 \\
  50--80\% & 2.84 $\pm$ 0.01 & 4.02 $\pm$ 0.01 \\
  \hline
  30--80\% & 3.48 $\pm$ 0.01 & 4.60 $\pm$ 0.01 \\
  \hline
\end{tabular}
\caption{Initial in-plane ($R_{x,\textrm{initial}}$) and
out-of-plane ($R_{y,\textrm{initial}}$) radii for 7 different
centrality bins.}\label{Rinitial}
\end{table}

For an azimuthally asymmetric collision, the initial source has an
elliptic shape with the larger axis perpendicular to the reaction
plane (out-of-plane) and the shorter axis in the reaction plane
(in-plane). In order to calculate the radii of the initial source
in the $x$ (in-plane) and $y$ (out-of-plane) direction we first
get the initial distribution of particles in the almond shaped
initial overlap from a Monte Carlo Glauber model calculation as
described in \cite{ADAMS03b}. The in-plane ($R_{x,initial}$) and
out-of-plane ($R_{y,initial}$) initial radii are calculated as the
radii of the region that contains 95\% of the particles. The
values for the initial in-plane and out-of-plane edge radii are
shown in Table \ref{Rinitial}. The azimuthally integrated initial
radius ($R_{\textrm{initial}}$) can be calculated from those two
radii as:
\begin{equation}
R_{\textrm{initial}}=\sqrt{\frac{R_{x,\textrm{initial}}^2+R_{y,\textrm{initial}}^2}{2}}.
\end{equation}

\begin{figure}
\includegraphics[width=0.50\textwidth]{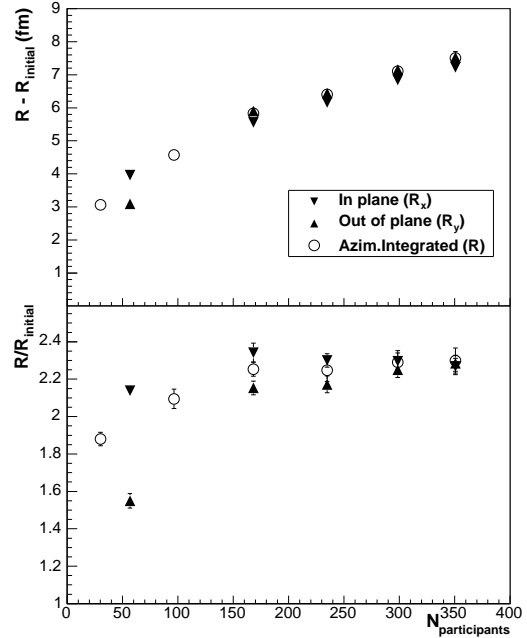}
\caption{$R-R_{\textrm{initial}}$ (top panel) and
$R/R_{\textrm{initial}}$ (bottom panel) for the azimuthally
integrated analysis and in the \textit{x} (in-plane) and
\textit{y} (out-of-plane) directions for the azimuthally-sensitive
case vs. number of participants.}\label{RsOMRMS}
\end{figure}

Figure \ref{RsOMRMS} (bottom panel) shows $R/R_{\textrm{initial}}$
vs. number of participants for in-plane, out-of-plane and
azimuthally integrated directions. The final source radii are
those extracted from the blast wave parametrization.
$R/R_{\textrm{initial}}$ is the relative expansion of the source
which is stronger in-plane than out-of-plane for the most
peripheral collisions, and it is similar in both directions for
the most central collisions. The azimuthally integrated radius
indicates a strong relative expansion of the source for central
collisions. This expansion seems to be very similar for all
centralities, decreasing just for the most peripheral cases.
Figure \ref{RsOMRMS} (top panel) shows the overall expansion of
the source given by $R-R_{\textrm{initial}}$ vs. number of
participants.

\begin{figure}
\includegraphics[width=0.50\textwidth]{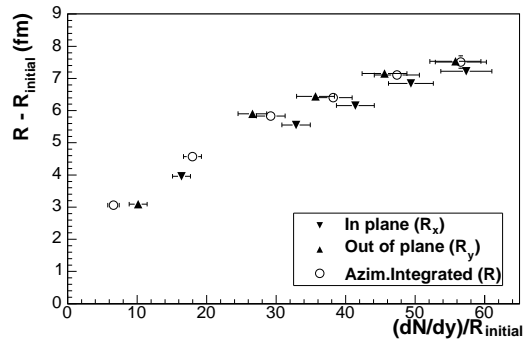}
\caption{$R-R_{\textrm{initial}}$ for the azimuthally integrated
analysis and in the \textit{x} (in-plane) and \textit{y}
(out-of-plane) directions for the azimuthally-sensitive case vs.
$(dN/dy)/R_{\textrm{initial}}$.}\label{RsmRivsdNdy}
\end{figure}

While the absolute expansion ($R-R_{\textrm{initial}}$) increases
steadily going to more central collision, the relative expansion
saturates when the number of participant reaches 150. Furthermore,
both absolute and relative expansions differ significantly in
peripheral events when comparing the in-plane and out-of-plane
directions, while they are similar for the most central. This is
expected arguing that the expansion, or in other word flow, is
driven by particle reinteractions following the initial pressure
gradients which in turn follow the initial energy density
gradients. As the centrality increases the difference between the
initial energy density gradient in-plane and out-of-plane
diminishes, which brings the expansions in-plane and out-of-plane
closer together.

The question is then what drives the transverse expansion. The
difference between the in-plane and out-of-plane expansion at a
given centrality shows that the initial energy density gradient
matters. The initial energy density gradients are responsible for
establishing the initial expansion velocity but the spatial
expansion will also depend on for how long the system expands. The
system lifetime is likely to depend on the initial energy density,
which may be gauged by dividing the particle multiplicity
($dN/dy$) by the initial area estimated in the Glauber framework
as done in \cite{ADAMS03b}. Following this idea, we investigate
how the transverse expansion evolves while varying centrality,
which affects both the average energy density and the energy
density gradient. We find that the transverse expansion scales
with $(dN/dy)/R_{\textrm{initial}}$ as shown in
Fig.~\ref{RsmRivsdNdy}. This figure shows $R -
R_{\textrm{initial}}$ vs. $(dN/dy)/R_{\textrm{initial}}$, where
$dN/dy$ is for pions as reported in \cite{ADAMS03} and
$R_{\textrm{initial}}$ is the corresponding in-plane, out-of-plane
or azimuthally integrated initial radius described above. This
quantity scales neither as a gradient nor as an energy density but
it appears to contain the relevant parameters that drive the
transverse expansion. We observe a clear scaling for $R_x$ and
$R_y$ as well as for the azimuthally integrated radius $R$, with
$(dN/dy)/R_{\textrm{initial}}$. For the same collisions, the
in-plane expansion corresponds to a higher value of
$(dN/dy)/R_{\textrm{initial}}$ than the corresponding out-of-plane
expansion.

The good fit to the data obtained with the blast wave
parametrization, consistent with expansion, and the comparison in
different ways of the initial and final sizes of the source
clearly indicate that the results can be interpreted in terms of
collective expansion that could be driven by the initial pressure
gradient. However, the time scales extracted from the fit seem to
be very small, smaller than the values predicted by hydrodynamic
models.

\begin{figure}
\includegraphics[width=0.50\textwidth]{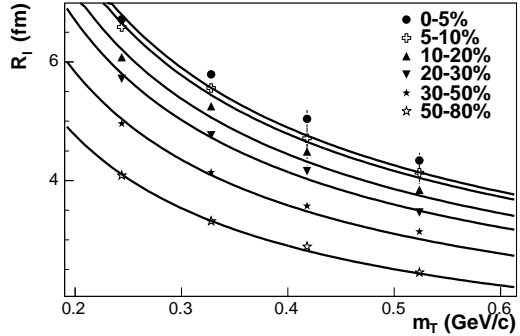}
\caption{Longitudinal HBT radius $R_l$. Lines represent the fits
$R_{l} = \tau
\sqrt{\frac{T}{m_{T}}\frac{K_{2}(m_{T}/T)}{K_{1}(m_{T}/T)}}$ for
each centrality.}\label{RlFit}
\end{figure}

\begin{figure}
\includegraphics[width=0.50\textwidth]{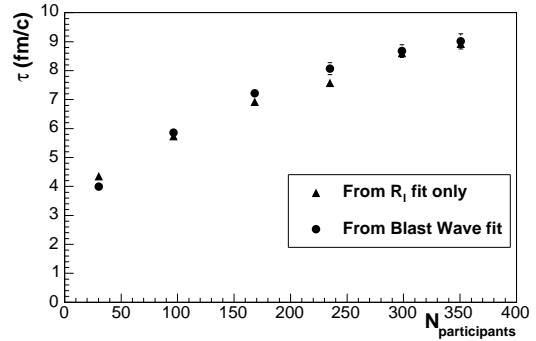}
\caption{Evolution time $\tau$ vs. number of participants as
extracted from a fit to $R_l$, lines in Figure \ref{RlFit}
(triangles), and from a blast wave fit to HBT parameters and
spectra (circles).}\label{Tau}
\end{figure}

From the dependence of $R_l$ on $m_T$ shown in Fig.~\ref{RlFit},
and assuming boost-invariant longitudinal flow, we can extract
information about the evolution time-scale of the source, or
proper time of freeze-out, by fitting it to a formula first
suggested by Sinyukov and collaborators \cite{MAKHL88,AKKEL95} and
then improved by others \cite{RETIE03}:
\begin{equation}
R_{l} =
\tau\sqrt{\frac{T}{m_{T}}\frac{K_{2}(m_{T}/T)}{K_{1}(m_{T}/T)}}
\end{equation}
where $T$ is the freeze-out temperature and $K_{1}$ and $K_{2}$
are the modified Bessel functions of order 1 and 2. This
expression for $R_l$ also assumes vanishing transverse flow and
instantaneous freeze-out in proper time (\textrm{i.e.},
$\Delta\tau$ = 0). The first assumption is approximatively
justified by the small dependence of $R_l$ on $\rho_0$ in full
calculation \cite{RETIE03}. The second approximation is justified
by the small $\Delta\tau$ from blast wave fits, Table
\ref{BWtable1}. Figure \ref{RlFit} also shows the fits to $R_l$
(lines) using temperatures, $T$, consistent with spectra as for
the fit to $R_s$. The extracted values for the evolution time
$\tau$ are shown in Fig.~\ref{Tau}. The evolution time increases
with centrality from $\tau \approx$ 4 fm/c for the most peripheral
events to $\tau \approx$ 9 fm/c for the most central events. In
the same plot, the extracted evolution time from the blast wave
fit is shown. Good agreement is observed between the two extracted
proper times for all centralities. They are surprisingly small as
compared with hydrodynamical calculations that predict a
freeze-out time of $\sim$~15 fm/c in central collisions. These
hydrodynamical calculations may over-predict the system lifetime
or the assumption on which the extraction of $\tau$ is based in
the blast wave parametrization, longitudinal boost invariant
expansion, might not be completely justified.

\begin{figure}
\includegraphics[width=0.50\textwidth]{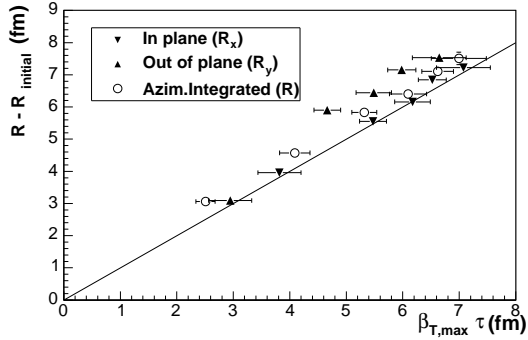}
\caption{$R-R_{\textrm{initial}}$ for the azimuthally integrated
analysis and for the in-plane and out-of-plane directions vs.
$\beta_{T,\textrm{max}}\cdot\tau$. The line is a ``$y=x$" line.}
\label{RsmRivsBetat}
\end{figure}

As a check for the consistency of the evolution time extracted
from the blast wave fit, Fig.~\ref{RsmRivsBetat} shows the final
source radius as extracted from the blast wave fit minus the
initial source size vs. $\beta_{T,\textrm{max}}\cdot\tau$. This
$\beta_{T,\textrm{max}}$ is the maximum flow velocity and is
expected to be the velocity at the edge of the expanding source at
kinetic freeze-out. It has been calculated from the $\rho_0$ and
$\rho_a$ blast wave parameters as $\beta_{T,\textrm{max}} =
\tanh(\rho_0+\rho_a\cos(2\phi))$. $\phi$ is 0 in-plane and $\pi/2$
out-of-plane \cite{RETIE03}, and $\rho_a$ is 0 for the azimuthally
integrated analysis and is given in Table \ref{BWtable2} for the
azimuthally sensitive analysis. The evolution time, $\tau$, is the
blast wave parameter shown in Fig.~\ref{Tau}, and Table
\ref{BWtable1}. The systematic errors in
$\beta_{T,\textrm{max}}\cdot\tau$ come from the finite size bin in
centrality. If the extracted radius and proper-time are right, the
initial and final edge radii should be related by the relation
$R_{\textrm{final}}~<~R_{\textrm{initial}}~+~\beta_{T,\textrm{max}}\cdot\tau$
so that the points in the figure should all be clearly below the
solid line
($\beta_{T,\textrm{max}}\cdot\tau~=~R-R_{\textrm{initial}}$).
Since most points are above the line, a possible explanation is
that $\tau$ is not properly calculated within the blast wave
parametrization. A larger $\tau$ would move the points below the
line.

\begin{figure}
\includegraphics[width=0.50\textwidth]{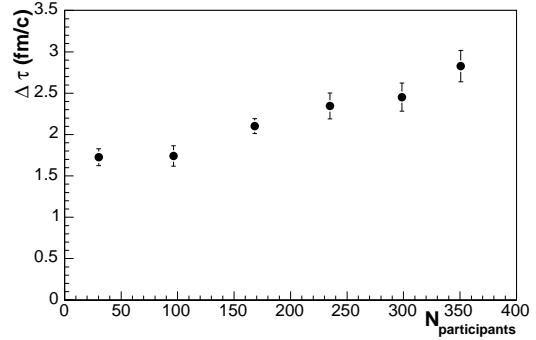}
\caption{Emission duration time $\Delta\tau$ vs. number of
participants as extracted using a blast fit to HBT parameters and
spectra.}\label{Deltat}
\end{figure}

Figure \ref{Deltat} shows the emission duration time, $\Delta\tau$
as a function of number of participants. $\Delta\tau$ increases
with increasing centrality up to $\sim$~3 fm/c. It is relatively
small for all centralities, however it has increased with respect
to the values extracted from our analysis at $\sqrt{s_{NN}}$ = 130
GeV \cite{RETIE03} due to the improved procedure of taking Coulomb
interaction into account and the consequent increase in $R_o$.

\begin{figure}
\includegraphics[width=0.50\textwidth]{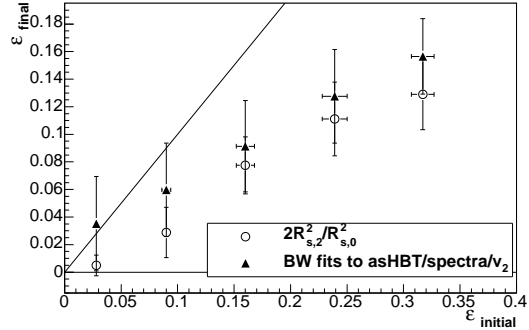}
\caption{Final source eccentricity
($\varepsilon_{\textrm{final}}$) as calculated from the Fourier
coefficients ($2R_{s,2}^2/R_{s,0}^2$) and from the final in-plane
and out-of-plane radii ($(R_y^2-R_x^2)/(R_y^2+R_x^2)$) vs. initial
eccentricity ($\varepsilon_{\textrm{initial}}$). The most
peripheral collisions correspond to the largest eccentricity. The
line indicates $\varepsilon_{\textrm{final}} =
\varepsilon_{\textrm{initial}}$. Systematic errors of 30\%, based
on sensitivity to model parameters \cite{RETIE03}, are assigned to
$\varepsilon_{\textrm{final}}$ extracted from the Fourier
coefficients.}\label{Eccent}
\end{figure}

The freeze-out shape of the source in non-central collisions and
its relation to the spatial anisotropy of the collision's initial
overlap region give us another hint about the system lifetime. The
initial anisotropic collision geometry generates greater
transverse pressure gradients in the reaction plane than
perpendicular to it. This leads to a preferential in-plane
expansion \cite{KOLB00,VOLOS03,ADLER01b} which diminishes the
initial anisotropy. A long-$\tau$ source would be less
out-of-plane extended and perhaps in-plane extended. The
eccentricity of the initial overlap region has been calculated
from the initial RMS of the distribution of particles, as given by
the Monte Carlo Glauber calculation, in the in-plane and
out-of-plane directions as:
\begin{equation}
\varepsilon =
\frac{(R_{y,\textrm{initial}}^{\textrm{RMS}})^2-(R_{x,\textrm{initial}}^{\textrm{RMS}})^2}
{(R_{y,\textrm{initial}}^{\textrm{RMS}})^2+(R_{x,\textrm{initial}}^{\textrm{RMS}})^2}.
\end{equation}

Figure \ref{Eccent} shows the relation between initial and final
eccentricities, with more peripheral collisions showing a larger
final anisotropy. The final source eccentricity has been
calculated from the Fourier coefficients ($2R_{s,2}^2/R_{s,0}^2$),
as well as from the final in-plane and out-of-plane radii
($(R_y^2-R_x^2)/(R_y^2+R_x^2)$) extracted from the blast wave fit
to azimuthally sensitive HBT and spectra described above. The
source at freeze-out remains out-of-plane extended indicating that
the outward pressure and/or expansion time was not sufficient to
quench or reverse the initial spatial anisotropy. The large
elliptic flow and small HBT radii observed at RHIC energies might
favor a large pressure build-up in a short-lived system compared
to hydrodynamic calculations. Also, out-of-plane freeze-out shapes
tend to disfavor a long-lived hadronic rescattering phase
following hydrodynamic expansion \cite{TEANE01}. This short
hadronic phase is consistent with a short emission time.

\section{Conclusion}\label{Concl}

We have presented a detailed description of a systematic HBT
analysis in Au+Au collisions at $\sqrt{s_{NN}}$ = 200 GeV. We have
analyzed the Gaussianess of the correlation function and conclude
that there is a deviation from a pure Gaussian, however it is not
clear how this affects the HBT parameters, extracted assuming a
Gaussian, that will be compared to models. We have studied the
centrality dependence of the $k_T$ dependence of the HBT
parameters and extracted geometrical and dynamical information on
the source at freeze-out. We conclude that there is a significant
expansion in Au+Au collisions, and that the relative expansion
does not significantly depend on centrality. The system expands by
a factor of at least 2.0 for most centralities. This is well
established by HBT. The initial pressure gradient seems to be
driving the expansion. The extracted time scales from a blast wave
fit are small. The blast wave evolution time $\tau$ is small as
compared with hydrodynamical calculations which could suggest that
the longitudinal boost invariant assumption has only limited
validity.

We thank the RHIC Operations Group and RCF at BNL, and the NERSC
Center at LBNL for their support. This work was supported in part
by the HENP Divisions of the Office of Science of the U.S. DOE;
the U.S. NSF; the BMBF of Germany; IN2P3, RA, RPL, and EMN of
France; EPSRC of the United Kingdom; FAPESP of Brazil; the Russian
Ministry of Science and Technology; the Ministry of Education and
the NNSFC of China; Grant Agency of the Czech Republic, FOM and UU
of the Netherlands, DAE, DST, and CSIR of the Government of India;
Swiss NSF; and the Polish State Committee for Scientific Research.

\end{document}

%% file: sci-jan04.tex
\affiliation{Argonne National Laboratory, Argonne, Illinois 60439}
\affiliation{University of Bern, 3012 Bern, Switzerland}
\affiliation{University of Birmingham, Birmingham, United Kingdom}
\affiliation{Brookhaven National Laboratory, Upton, New York
11973} \affiliation{California Institute of Technology, Pasedena,
California 91125} \affiliation{University of California, Berkeley,
California 94720} \affiliation{University of California, Davis,
California 95616} \affiliation{University of California, Los
Angeles, California 90095} \affiliation{Carnegie Mellon
University, Pittsburgh, Pennsylvania 15213} \affiliation{Creighton
University, Omaha, Nebraska 68178} \affiliation{Nuclear Physics
Institute AS CR, 250 68 \v{R}e\v{z}/Prague, Czech Republic}
\affiliation{Laboratory for High Energy (JINR), Dubna, Russia}
\affiliation{Particle Physics Laboratory (JINR), Dubna, Russia}
\affiliation{University of Frankfurt, Frankfurt, Germany}
\affiliation{Insitute  of Physics, Bhubaneswar 751005, India}
\affiliation{Indian Institute of Technology, Mumbai, India}
\affiliation{Indiana University, Bloomington, Indiana 47408}
\affiliation{Institut de Recherches Subatomiques, Strasbourg,
France} \affiliation{University of Jammu, Jammu 180001, India}
\affiliation{Kent State University, Kent, Ohio 44242}
\affiliation{Lawrence Berkeley National Laboratory, Berkeley,
California 94720} \affiliation{Massachusetts Institute of
Technology, Cambridge, MA 02139-4307}
\affiliation{Max-Planck-Institut f\"ur Physik, Munich, Germany}
\affiliation{Michigan State University, East Lansing, Michigan
48824} \affiliation{Moscow Engineering Physics Institute, Moscow
Russia} \affiliation{City College of New York, New York City, New
York 10031} \affiliation{NIKHEF, Amsterdam, The Netherlands}
\affiliation{Ohio State University, Columbus, Ohio 43210}
\affiliation{Panjab University, Chandigarh 160014, India}
\affiliation{Pennsylvania State University, University Park,
Pennsylvania 16802} \affiliation{Institute of High Energy Physics,
Protvino, Russia} \affiliation{Purdue University, West Lafayette,
Indiana 47907} \affiliation{University of Rajasthan, Jaipur
302004, India} \affiliation{Rice University, Houston, Texas 77251}
\affiliation{Universidade de Sao Paulo, Sao Paulo, Brazil}
\affiliation{University of Science \& Technology of China, Anhui
230027, China} \affiliation{Shanghai Institute of Applied Physics,
Shanghai 201800, China} \affiliation{SUBATECH, Nantes, France}
\affiliation{Texas A\&M University, College Station, Texas 77843}
\affiliation{University of Texas, Austin, Texas 78712}
\affiliation{Tsinghua University, Beijing 100084, China}
\affiliation{Valparaiso University, Valparaiso, Indiana 46383}
\affiliation{Variable Energy Cyclotron Centre, Kolkata 700064,
India} \affiliation{Warsaw University of Technology, Warsaw,
Poland} \affiliation{University of Washington, Seattle, Washington
98195} \affiliation{Wayne State University, Detroit, Michigan
48201} \affiliation{Institute of Particle Physics, CCNU (HZNU),
Wuhan 430079, China} \affiliation{Yale University, New Haven,
Connecticut 06520} \affiliation{University of Zagreb, Zagreb,
HR-10002, Croatia}

\author{J.~Adams}\affiliation{University of Birmingham, Birmingham, United Kingdom}
\author{M.M.~Aggarwal}\affiliation{Panjab University, Chandigarh 160014, India}
\author{Z.~Ahammed}\affiliation{Variable Energy Cyclotron Centre, Kolkata 700064, India}
\author{J.~Amonett}\affiliation{Kent State University, Kent, Ohio 44242}
\author{B.D.~Anderson}\affiliation{Kent State University, Kent, Ohio 44242}
\author{D.~Arkhipkin}\affiliation{Particle Physics Laboratory (JINR), Dubna, Russia}
\author{G.S.~Averichev}\affiliation{Laboratory for High Energy (JINR), Dubna, Russia}
\author{S.K.~Badyal}\affiliation{University of Jammu, Jammu 180001, India}
\author{Y.~Bai}\affiliation{NIKHEF, Amsterdam, The Netherlands}
\author{J.~Balewski}\affiliation{Indiana University, Bloomington, Indiana 47408}
% ...

\collaboration{STAR Collaboration}\noaffiliation
